\documentclass[a4paper,11pt]{article}
\pdfoutput=1 

\usepackage{jcappub} 

\usepackage[T1]{fontenc} 
\usepackage{graphicx}
\usepackage{caption}
\usepackage{floatrow}   
\usepackage{subcaption}
\usepackage{multirow}
\usepackage{xcolor}

\newcommand{\RefReport}[1]{{\color{black} {#1}}}
\newcommand{\RefReportTwo}[1]{{\color{black} {#1}}}

\title{\boldmath Emulation of baryonic effects on the matter power spectrum and constraints from galaxy cluster data}

\author[]{Sambit K. Giri}
\author[]{and Aurel Schneider}

\affiliation[]{Institute for Computational Science, University of Zurich, \\Winterthurerstrasse 190, 8057 Zurich, Switzerland}

\emailAdd{sambitkumar.giri@uzh.ch}
\emailAdd{aurel.schneider@uzh.ch}

\abstract{Baryonic feedback effects consist of a major systematic for upcoming weak-lensing and galaxy-clustering surveys. In this paper, we present an emulator for the baryonic suppression of the matter power spectrum. The emulator is based on the baryonification model, containing seven free parameters that are connected to the gas profiles and stellar abundances in haloes. We show that with the baryonic emulator, we can not only recover the power spectra of hydro-dynamical simulations at sub-percent precision, but also establish a connection between the baryonic suppression of the power spectrum and the gas and stellar fractions in haloes. This connection allows us to predict the expected deviation from a dark-matter-only power spectrum using measured X-ray gas fractions of galaxy groups and clusters. With these measurements, we constrain the suppression to exceed the percent-level at $k=0.1-0.4$ h/Mpc and to reach a maximum of 20-28 percent at around $k\sim 7$ h/Mpc (68 percent confidence level). As a further step, we also perform a detailed parameter study and we present a minimum set of four baryonic parameters that are required to recover the scale and redshift dependence observed in hydro-dynamical simulations. The baryonic emulator can be found at \url{https://github.com/sambit-giri/BCemu}. 
}


\begin{document}
\maketitle
\flushbottom

\section{Introduction}
\label{sec:intro}
The next generation of cosmological surveys will not only lead to a massive increase of data but also allow us to push towards smaller cosmological scales. While these scales may carry new information regarding the underlying cosmology, they are at the same time strongly nonlinear and affected by astrophysical phenomena, making them very challenging to model. One of the main concerns in that respect \RefReport{is} active galactic nuclei (AGN) expelling large amounts of gas into the intergalactic medium (IGM). Such AGN feedback processes cannot be modelled from first principles, and they have been shown to alter the clustering signal at scales relevant \RefReport{to} cosmology \cite{VanDaalen2011,semboloni2011quantifying,harnois2015baryons,huang2019modelling}.

Based on various cosmological simulations, it has been shown that baryonic (AGN) feedback leads to a suppression of the matter power spectrum of order 10 percent at non-linear scales, i.e. wave-modes beyond $k\sim0.3$ h/Mpc \cite{chisari2019modelling}.
While most simulations predict a similar general trend, they do not agree \RefReport{with} 
each other at the quantitative level. This should, however, not come as a surprise, since AGN feedback processes are included via different sub-grid models \RefReport{in} the simulations. On top of that, the model parameters are usually tuned to reproduce key galaxy properties without a specific focus on effects relevant \RefReport{to} cosmology.

The specific connection between AGN feedback parameters and cosmology was first investigated using the suites of hydro-dynamical simulations {\tt OWLS} \cite{schaye2010physics} 
and {\tt cosmo-OWLS} \cite{le2014towards,mccarthy2014thermal}. 
These simulations revealed that the strength of the feedback ejection has a direct influence on the matter power spectrum and on the amount of gas within haloes. The latter can be directly measured using X-ray observations of individual galaxy groups and clusters. As a consequence, the AGN sub-grid model parameters in simulations can be directly tuned to X-ray observations, a strategy pursued for the cosmological simulation suite {\tt BAHAMAS} \cite{Mccarthy2017TheCosmology,Mccarthy2018TheFormation}. 

Recently it has been shown that, instead of running expensive hydro-dynamical simulations, it is possible to include baryonic effects in gravity-only $N$-body simulation outputs by perturbatively modifying the positions of simulation particles around haloes \cite{Schneider2015ASpectrum}. This baryonification method is based on parametrised gas and stellar profiles with shapes motivated by observations. It comes with a handful of well-defined, physically meaningful baryonic parameters, which can be systematically varied in a cosmological context. Note furthermore, that the baryonification method has shown to provide predictions that are accurate enough for the next-generation cosmological surveys, at least at the level of the matter power spectrum \cite{Schneider2019QuantifyingCorrelation,Arico2021SimultaneousBaryons}.

Although the baryonification method is many orders of magnitude faster than running full hydro-dynamical simulations, it still requires the modification of large $N$-body output files, making it unsuitable to be used directly in Bayesian parameter inference pipelines. The main goal of the present paper is therefore to construct an emulator for the baryonic suppression effect on the matter power spectrum. This emulator is based on the same set of parameters that the underlying baryonification model and it allows for more efficient exploration of the vast parameter space. Since there is no hope of determining the effects of baryonic feedback from first principles, this is the only way of constraining baryonic nuisance parameters in a cosmological context.

After building the emulator based on a training set from a slightly modified version of the baryonification model of Ref. \cite{Schneider2019QuantifyingCorrelation}, we compare it to results from a variety of hydro-dynamical simulations available in the literature. The goal is to not only show that the simulated power spectra can be recovered at percent accuracy, but also to highlight that the baryonification method is able to predict the simulated power spectra based on information about the gas and stellar abundance in haloes. If both these aspects are confirmed, we can use the baryonification model for cross-correlation analyses~\cite{Schneider2019BaryonicMatrix}.

Other questions we want to address in the present paper are: what is the minimum amount of baryonic nuisance parameters that \RefReport{can} 
recover hydro-dynamical simulation results at acceptable accuracy? Can we reproduce the redshift dependence visible in hydro-dynamical simulations? What can we learn from X-ray observations regarding the expected baryonic suppression of the power spectrum? 

The paper is structured as follows: in Sec.~\ref{sec:brf_model} we provide a summary of the baryonification model and describe the set-up of the emulator and quantify its accuracy. Sec.~\ref{sec:hydro_sims} is dedicated to a comparison with hydro-dynamical simulations. In Sec.~\ref{sec:constraints_obs} we introduce a case study where we use the measured gas fractions of X-ray measurements of galaxy clusters to predict the power suppression due to baryonic feedback.
In Sec.~\ref{sec:reduce_param} we present a way to reduce the number of free model parameters, while maintaining an acceptable agreement with simulations. We next study the redshift dependence of the model parameters in Sec.~\ref{sec:zdep_params}. 
Finally we present our conclusions in Sec.~\ref{sec:conclusion}.

\section{Baryonification model and emulator}
\label{sec:brf_model}
The method of baryonifying $N$-body simulations is based on the principle of modifying halo profiles in simulation outputs by slightly displacing particles around halo centres. Although this displacement is performed in a spherically symmetric way, the resulting $N$-body outputs still contain the full complexity of the nonlinear cosmological density field with non-spherical haloes arranged along the cosmic web. The result of the baryonification process is primarily visible in statistical measures such as the matter power spectrum.

In this section we first summarise the important ingredients of the baryonification model with specific emphasis on the free model parameters. We then present the emulator of the baryonic power suppression. A specific emphasis is placed on the production of the data set, the emulation method, and the testing phase. 

\subsection{Model ingredients}
\label{sec:brf_model_theory}
The guiding principle of the baryonification method is the transformation of gravity-only profiles into modified, dark-matter-baryon (dmb) profiles that include the effects of dark matter, gas, and stars \citep{Schneider2015ASpectrum}. This process can be described as
\begin{equation}\label{rhodmb}
\rho_{\rm nfw}(r)\,\,\,\rightarrow\,\,\, \rho_{\rm dmb}(r) = \rho_{\rm clm}(r) + \rho_{\rm gas}(r) + \rho_{\rm cga}(r),
\end{equation}
where $\rho_{\rm nfw}$ denotes the truncated NFW profile \cite{Navarro1996nfw,Baltz2009AnalyticPotentials}, while $\rho_{\rm dmb}$ is composed of a collision-less matter ($\rho_{\rm clm}$), gas ($\rho_{\rm gas}$), and central galaxy profile ($\rho_{\rm cga}$). The collision-less profile mainly consists of dark matter but also includes all stars in satellite galaxies and the stellar halo.

The stellar profile of the central galaxy is described by an exponentially truncated power law
\begin{equation}\label{rhocga}
\rho_\mathrm{cga}(r) = \frac{f_\mathrm{cga}(M)}{4\pi^{3/2}R_\mathrm{h}r^2}  \exp \left[ -\left( \frac{r}{2R_\mathrm{h}} \right)^2 \right],
\end{equation}
where $f_{\rm cga}(M)$ describes the fraction of stars in the central galaxy (defined in Eq.~\ref{stellarfractions} below) and $R_\mathrm{h}$ is the stellar half-light radius which is set to 0.015 times the virial radius $r_\mathrm{vir}$ \cite[a value in agreement with observations, see e.g. Ref.][]{Kravtsov2018StellarHalos}. Note that for 
the matter power spectrum at cosmological scales, the details of the stellar profiles are not important and the simplistic profile of Eq.~(\ref{rhocga}) is therefore sufficient.

The distribution of gas in the halo is parametrised by a cored double-power law of the form
\begin{equation}\label{rhogas}
\rho_\mathrm{gas}(r) \propto \frac{\left[\Omega_b/\Omega_m-f_\mathrm{star}(M)\right]}{\left[1+ 10\left(\frac{r}{r_\mathrm{vir}}\right) \right]^{\beta(M)} \left[1+ \left(\frac{r}{\theta_\mathrm{ej}r_\mathrm{vir}}\right)^\gamma \right]^{\frac{\delta-\beta(M)}{\gamma}}},
\end{equation}
where $f_{\rm star}(M)$ is the total fraction of stars in haloes (see Eq.~\ref{stellarfractions} below). The $\beta$-parameter has an additional halo-mass dependence which is given by
\begin{equation}\label{eq:beta_model}
\beta(M_c,\mu) = \frac{3(M/M_c)^\mu}{1+(M/M_c)^\mu} \ ,
\end{equation}
allowing the gas profile to become shallower than the NFW profile towards smaller halo masses ($M\lesssim M_c$). Note, furthermore that the mass-dependence of $\beta(M)$ differs slightly from the definition in Ref.~\cite{Schneider2019BaryonicMatrix,Schneider2019QuantifyingCorrelation}, in the sense that it is now strictly limited to positive values \citep[see e.g. Ref.][]{Mead2021Hmcode-2020Feedback}.

The gas profile of Eqs.~(\ref{rhogas},
\ref{eq:beta_model}) contains the five free model parameters $M_c$, $\mu$, $\theta_{\rm ej}$, $\gamma$, and $\delta$. These parameters are crucial for the modelling of baryonic feedback effects, as they determine how the gas is distributed in and around haloes.

The collision-less matter profile is dominated by the dark matter component, but also contains satellite galaxies and intra-cluster stars. At first, it is given by
\begin{equation}
\rho_\mathrm{clm}(r) = \left[\frac{\Omega_\mathrm{dm}}{\Omega_\mathrm{m}} + f_\mathrm{sga}(M)\right] \rho_\mathrm{nfw}(r)
\end{equation}
where $\rho_{\rm nfw}$ is the truncated NFW profile and $f_{\rm sga}(M)$ is the stellar fraction attributed to satellite galaxies and the intra-cluster light.  Note, that the collision-less profile may react to the other components via the adiabatic relaxation prescription of Ref. \cite{Abadi2010Galaxy-inducedHaloes}. As a consequence, it does not retain its original NFW shape. A more detailed description of this procedure can be found in Ref. \cite{Schneider2019QuantifyingCorrelation}.

After having discussed the three components making up the dark-matter-baryon profile in Eq.~(\ref{rhodmb}), we will now introduce the stellar-to-halo fractions that appear as part of these components. The stellar fraction of satellites (and halo stars) is obtained by subtracting the central galactic fraction from the total fraction of stars in haloes, i.e.,
\begin{equation}
f_{\rm sga}(M)= f_{\rm star}(M) - f_{\rm cga}(M),
\end{equation}
where $M$ refers throughout the paper to the halo mass at 200 times the critical density. The total stellar fraction and the fraction of stars in the central galaxy are each given by a power law
\begin{equation}\label{stellarfractions}
 f_{i}(M) = 0.055 \left(\frac{M}{M_{s}} \right )^{-\eta_{i}}
\end{equation}
with $i=\lbrace {\rm star}, {\rm cga}\rbrace$ and $M_s=2.5\times 10^{11} h^{-1}M_\odot$. Note that this simple description is influenced by the well-known \cite{Moster2013} relation, but only contains the large-scale decrease (since haloes below $\sim10^{12}$ M$_{\odot}$ do not have to be modelled for our purposes). We furthermore redefine $\eta_{\rm star}\equiv\eta$ and $\eta_{\rm cga}=\eta+\eta_\delta$, ending up with 2 additional parameters ${\lbrace\eta,\eta_\delta\rbrace}$. All seven model parameters ${\lbrace M_c, \mu, \theta_{\rm ej}, \gamma, \delta, \eta,\eta_\delta\rbrace}$  describing the gas distribution and stellar abundances, are summarized in Table \ref{table:range_values}.

\subsection{Baryonic emulator}
After summarising the baryonification model, we now introduce the baryonic emulator for the matter power spectrum. We start by presenting the training set, before discussing the emulation and testing method.

\subsubsection{Dataset}



To simulate the distribution of dark matter in our Universe, we run an $N$-body simulation with $512^3$ particles in a cubic volume of $(256~ h^{-1} \mathrm{Mpc})^3$ using the code \texttt{Pkdgrav3} \cite{2017ComAC...4....2P}. This setup has been shown in Ref.~\cite{Schneider2015ASpectrum} to be sufficient to study \RefReport{the} baryonic suppression effects of the power spectrum at sub-percent accuracy. We assume a standard 5-parameter $\Lambda$CDM cosmology where $\Omega_m=0.315$, $\Omega_b=0.049$, $h_0=0.674$, $n_s=0.96$ and $\sigma_8=0.811$. This cosmology is within the constraints given by \texttt{Planck} \cite{PlanckCollaboration2018PlanckParameters}. Previous studies have shown that the baryonic effects are insensitive to (reasonable) changes \RefReport{in} 
cosmological parameter values except for the mean baryon fraction $f_b=\Omega_b/\Omega_m$ \cite{Schneider2019BaryonicMatrix,Arico2021TheNetworks}. Therefore, our baryonic emulator 
will only depend on cosmology via $f_b$ which greatly simplifies the analysis. 

\begin{table}[h!]
\centering
\begin{tabular}{c c} 
 Parameter name & Range  \\ 
 \hline \hline
 $\mathrm{log_{10}} M_\mathrm{c}$ & [11, 15]\\ 
 $\mu$ & [0, 2] \\
 $\theta_\mathrm{ej}$ & [2, 8] \\
 $\gamma$ & [1, 4] \\
 $\delta$ & [3, 11] \\
 $\eta$ & [0.05, 0.4] \\
 $\eta_{\delta}$ & [0.05, 0.4] \\
 $f_b$ & [0.10, 0.25] \\
\end{tabular}
\caption{List of parameter abbreviations and ranges used for the baryonic emulator.  The top five parameters ($\log_\mathrm{10} M_c$, $\mu$, $\theta_{\rm ej}$, $\gamma$, $\delta$) are related to the gas profiles, parameters 6 and 7 determine the stellar fractions ($\eta$, $\eta_\delta$), while the bottom parameter ($f_b$) describes the relevant changes in cosmology.}
\label{table:range_values}
\end{table}

We create the training set by varying seven baryonification parameters along with $f_b$.
In Table~\ref{table:range_values}, we list the ranges of these parameters $\pmb{\theta}_\mathrm{bar}$ on which our emulator will depend as well as their respective ranges. This eight-parameter space is randomly sampled with 2700 points. 
\RefReport{We discuss our procedure for constructing the training set in Appendix~\ref{sec:const_training_set}}.
We then apply our baryonification algorithm at each point before measuring the power spectra of the 2700 modified $N$-body outputs. The power spectra are calculated between wave-modes $0.03~\mathrm{h/Mpc} \lesssim k \lesssim 12.52~\mathrm{h/Mpc}$ in 232 bins.
We finally divide each absolute power spectrum by the dark-matter-only power spectrum of the original simulation output in order to obtain the relative baryonic power suppression ($\pmb{\mathcal{S}}$). This training set is used to create the emulator (described in later sections).

Next to the training set, we  independently create another data set with 300 randomly selected points in our eight-dimensional parameter space. This corresponds to the testing set that will be used to estimate the accuracy of our emulator.

\subsubsection{Method}
We build our emulator by fitting a regression model to our data-set $\{ \Tilde{\pmb{\theta}}_\mathrm{bar}, \Tilde{\pmb{\mathcal{S}}}\}$, where $\Tilde{\pmb{\theta}}_\mathrm{bar}$ and $\Tilde{\pmb{\mathcal{S}}}$ are baryonification parameter and power spectrum suppression vector, respectively. We use \textit{kriging}, which is an interpolation based on \RefReport{the} Gaussian process (GP), to determine the regression model. In the most generic form, the kriging model can be written as \cite[e.g.][]{roustant2012dicekriging},
\begin{eqnarray}
 \pmb{\mathcal{S}} = \sum_{i=1}^{p}f_i(\pmb{\theta}_\mathrm{bar}) + \mathcal{Z}(\pmb{\theta}_\mathrm{bar}) \ ,
\end{eqnarray}
where the first term is a linear combination of a known function $f_i$ and the second term is a realisation of a stochastic process or GP, $\mathcal{Z}(\pmb{\theta}_\mathrm{bar})\sim \mathcal{N}(0,\mathcal{K})$. We assume a linear polynomial for $f_i$.
$\mathcal{K}$ is the kernel that quantifies the similarity between the points in the data set. In this work, we use the squared-exponential kernel defined as,
\begin{eqnarray}
 \mathcal{K}(\pmb{\theta}_\mathrm{bar},\pmb{\theta}_\mathrm{bar}^\prime) = \sigma^2 \prod_{l=1}^{d} \exp \left(-A_l({\theta}_{\mathrm{bar},l}-{\theta}_{\mathrm{bar},l}^\prime)^2\right) \ ,
 \label{eq:kernel_SEK}
\end{eqnarray}
where $d$ is the dimension of each data point in $\pmb{\theta}_\mathrm{bar}$. 
$A_l$'s and $\sigma$ are the hyper-parameters  that are determined during training. Once the hyper-parameters are known,  Eq.~\ref{eq:kernel_SEK} can be solved at any new point of the parameter space. However, this calculation can be computationally expensive especially for large data sets. We therefore use a modified version of kriging, which is known as \textit{kernel partial least squares} (KPLS) \cite{bouhlel2016improving}. 
Emulators built with KPLS are much faster and require less memory without loss in accuracy compared to the standard kriging method\footnote{This speed-up is obtained  due to \RefReport{the} inclusion of \RefReport{the} partial least square (PLS) method \cite{wold1975soft}. The PLS method is a tool for high dimensional problems that searches the direction that maximises the variance between input and output variables. The PLS information is integrated into the kriging correlation matrix to scale down computational complexity. \RefReport{Note that in previous works, principal component analysis (PCA) has been used to reduce the data before training an emulator using the kriging method \cite[e.g.][]{angulo2020bacco,Arico2021TheNetworks,parimbelli2021mixed}. The advantage of using PLS while fitting a regression model is that it maintains the correlation between the dependent and independent variables \cite{maitra2008principle}. However, a direct comparison between emulators built on data reduced with PLS and PCA methods is beyond the scope of this paper.}}. 
This makes them more suitable for Bayesian inference pipelines. We use the KPLS implementation in the \texttt{SMT}\footnote{This package is available at \url{https://smt.readthedocs.io}.} \cite{SMT2019}. 

In this work, we build independent emulators at each redshift ($z=0,0.5,1,1.5,2$). Note that Ref.~\cite{Arico2021TheNetworks} took a different approach by adding $z$ as an additional emulation parameter. We follow the strategy of multiple emulators at different redshifts for two main reasons: First, this approach allows us to implement different models for the \RefReport{redshift} evolution of individual parameters after the emulators are constructed. Indeed, any redshift evolution of individual parameters can be trivially \RefReport{extracted} from the emulators at different redshifts as long as the parameter values (that are now evolved with redshift) remain in the range of the emulator. Second, not emulating redshift allows us to reduce the dimension of the regression model. Due to the \textit{curse of dimensionality}, the required size of the training set increases substantially due to increase in the dimension \cite{Bellman2003DynamicProgramming,Theodoridis2008PatternRecognition}.

In Sec.~\ref{sec:reduce_param}, we will specifically study the redshift evolution of emulation parameters. More particularly, we investigate what parameters need to change with redshift in order to reproduce the observed evolution of the baryonic power suppression in hydro-dynamical simulation.

Our baryonic emulator is available online and can be used by installing \texttt{BCemu}\footnote{This package is available at \url{https://github.com/sambit-giri/BCemu}.}. Note that this emulator not only predicts the power spectrum suppression for a given set of baryonic parameters but 
also provides the uncertainty \RefReport{due to interpolation.}

\subsubsection{Testing emulation accuracy}

 \begin{figure}[t] 
 \centering
  \includegraphics[width=1.0\textwidth]{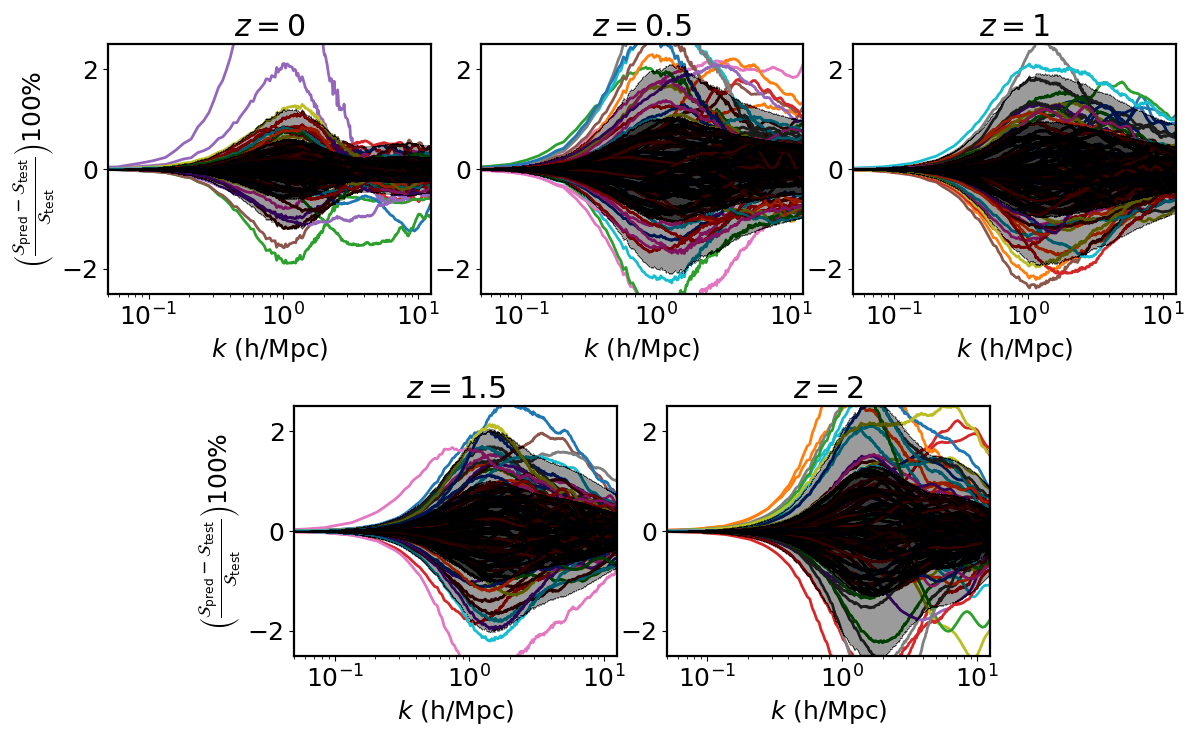}
 \caption{The percentage error of predicting the $\mathcal{S}(k)$ in the testing set with our emulators at various redshifts \RefReport{are shown using the coloured lines.}. The dark and light shaded regions represent the 1-$\sigma$ and 2-$\sigma$ regions respectively.}
 	\label{fig:error_percent_testset}
 \end{figure}

In this section, we investigate the performance of the emulator using the testing set.
We quantify the accuracy of our emulator with the coefficient of determination (also known as $R^2$ score) given as,
\begin{eqnarray}
 R^2 = \sum_{j=1}^{n_\mathrm{bins}} \left[1- \frac{\sum_{i=1}^{n_\mathrm{test}} \left(y_i(k_j)-\hat{y}_i(k_j)\right)^2}{\sum_{i=1}^{n_\mathrm{test}} \left(y_i(k_j)-\bar{y}(k_j)\right)^2} \right] \ ,
 \label{eq:r2_full}
\end{eqnarray}
where $\bar{y}(k_j) = \frac{1}{n_\mathrm{test}} \sum_{i=1}^{n_\mathrm{test}} y_i(k_j)$. $n_\mathrm{test}$ and $n_\mathrm{bins}$ are the size of the testing set and \RefReport{the} number of $k$ bins, respectively. The $R^2$ score measures 
how well the model or emulator replicates the outcomes $y$.
In order to determine the accuracy of our emulator in predicting the testing set, $y_i$ is replaced with $\mathcal{S}_i(k)$ in the above equation. We find that the $R^2$ score for the testing set is more than 0.99 at \RefReport{$z\lesssim 1.5$}, 
which corresponds to a very high accuracy level. \RefReport{The corresponding score at $z=2$ is 0.975.}

In Figure \ref{fig:error_percent_testset}, we show the relative percentage error between the power spectra suppression in the testing set $\mathcal{S}_\mathrm{test}$ and the corresponding predicted power spectra suppression $\mathcal{S}_\mathrm{pred}$. The dark-black and light-black regions correspond to the 1$\sigma$ and 2$\sigma$ regions, respectively.
\RefReport{The 1$\sigma$ regions stay within 1 percent when compared to the true answer at all wave-modes and redshifts. The 2$\sigma$ regions stay within 2 percent for $z<2$ and within 3 percent 
for $z=2$ at all wave-modes. In general, we note that the performance of the emulator becomes slightly worse towards higher redshifts. We speculate that this is a consequence of the fact that the prior ranges of the baryonic parameters are selected based on X-ray observations from low redshifts. It is possible that the same prior ranges lead to more variability of the power spectrum at higher redshifts, making the emulation of the suppression signal somewhat more challenging.}

\section{Studying baryonic effects in hydro-dynamical simulations}
\label{sec:hydro_sims}

In this section we test the validity of the baryonification method  by comparing it to hydro-dynamical simulations. This comparison is important not only to assure that the matter power spectrum can be reproduced at percent level but also to investigate the connection between halo properties and the power spectrum. The latter is a prerequisite for combining cosmological data with individual observations of gas around haloes, which consists of a potentially powerful way to constrain baryonic feedback properties \cite{Schneider2020wlxray}.

In this paper we consider the hydro-dynamical simulations {\tt OWLS} \cite{schaye2010physics}, {\tt cosmo-OWLS} \cite{le2014towards}, {\tt BAHAMAS} \cite{Mccarthy2017TheCosmology,Mccarthy2018TheFormation}, {\tt Illustris-TNG} \cite{springel2018first,weinberger2016simulating,pillepich2018simulating,pillepich2018StellarMass}, and {\tt Horizon-AGN} \cite{chisari2018impact}. The reason for selecting these simulations is that we could find published gas and stellar fractions next to the matter power spectrum. This is a requirement for the type of comparison analysis carried out below. A list of the simulations used for the comparison is provided in Table~\ref{table:hydro_sims}.
 
\begin{table}[h!]
\centering
\begin{tabular}{c c c c c} 
 \multirow{2}{4em}{Simulation \\ \ \ \ name} & cosmology & code & AGN feedback & \multirow{2}{4em}{box size ($h^{-1}$Mpc)} \\ \\
 \hline \hline
 {\tt BAHAMAS} \cite{Mccarthy2017TheCosmology,Mccarthy2018TheFormation} & {\tt WMAP9} \cite{hinshaw2013nine} & {\tt Gadget3} \cite{spergel2007three} & thermal & 400 \\ 
 {\tt cosmo-OWLS} \cite{le2014towards} & {\tt WMAP7} \cite{Komatsu2011Seven-yearInterpretation} & {\tt Gadget3} \cite{spergel2007three} & thermal & 400 \\ 
 {\tt OWLS} \cite{schaye2010physics,semboloni2011quantifying} & {\tt WMAP3} \cite{spergel2007three} & {\tt Gadget3} \cite{spergel2007three} & thermal & 100 \\ 
 {\tt Illustris-TNG} \cite{springel2018first,weinberger2016simulating,pillepich2018simulating,pillepich2018StellarMass} & {\tt Planck} \cite{planck2016CosmologicalParameters} & {\tt AREPO} \cite{springel2010pur} & thermal and kinetic & 75 \\ 
 {\tt Horizon-AGN} \cite{chisari2018impact} & {\tt WMAP7} \cite{Komatsu2011Seven-yearInterpretation} & {\tt RAMSES} \cite{Teyssier2002RAMSES} & thermal and kinetic & 100 \\ 
\end{tabular}
\caption{List of hydro-dynamical simulations considered in this work along with the cosmology, the simulation code, and the implemented AGN feedback mode.}
\label{table:hydro_sims}
\end{table}

\subsection{Reproducing the power suppression signal from simulations}
The first and most important requirement of our baryonification model is its ability to reproduce the baryonic power suppression predicted by simulations. We use the 7-parameter baryonification model to fit the emulated relative power spectra $\mathcal{S}$ to the results from the hydro-dynamical simulations. The best fits are obtained using the maximum likelihood estimation (MLE) technique\footnote{We use the optimization methods implemented in {\tt scipy} \cite{virtanen2020scipy} python package to maximize the likelihood.} \cite{chambers2012maximum,rossi2018mathematical}. For the likelihood, we assume a Gaussian error of 1 percent at each wave-mode of the $\mathcal{S}(k)$ we want to fit.

\begin{figure}[t] 
 \centering
\includegraphics[width=0.95\textwidth]{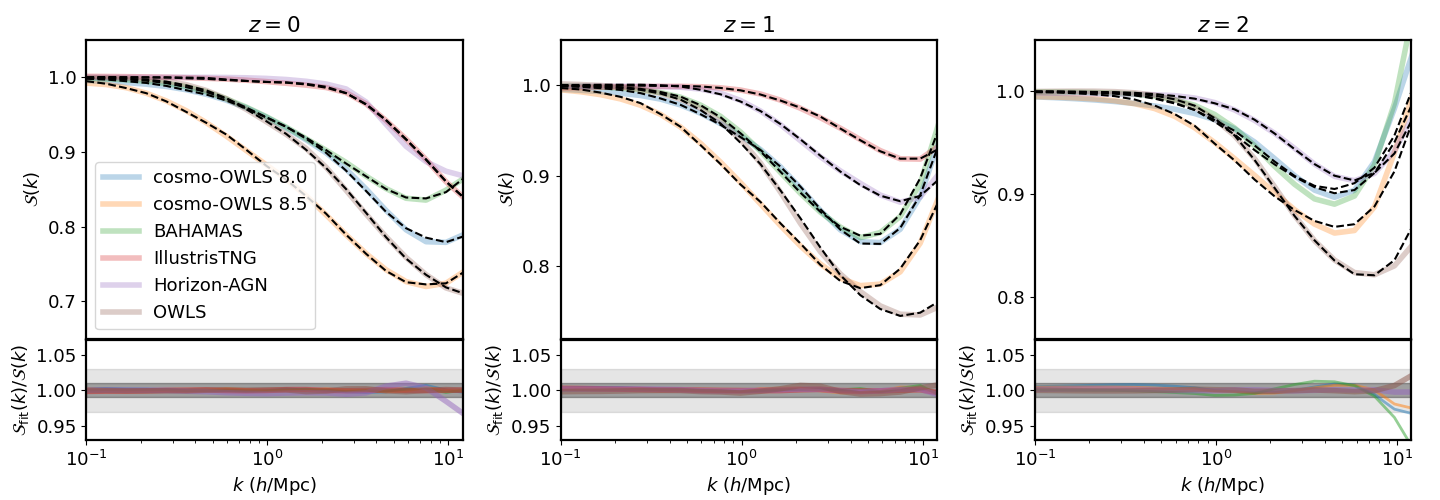}
 \caption{The relative matter power spectrum $\mathcal{S}=P/P_{\rm N-body}$ from all the hydro-dynamical simulations along with the best fitting curves from the baryonification model (black-dashed lines) at three redshifts. The agreement between model and simulations remains better than one percent over all $k$-modes considered (see bottom panel) at low redshifts ($z<2$).
 At $z=2$, the agreement degrades at small scales ($k\gtrsim 8$ h/Mpc).
 }
\label{fig:best_fit_7param}
\end{figure}

The best-fitting curves together with the results from the different hydro-dynamical simulations are shown in Fig.~\ref{fig:best_fit_7param}. The agreement between hydro-dynamical simulations and the baryonification model is better than one percent \RefReport{at all the} wave-modes considered at low redshifts ($z<2$). 
Only for the $z=2$ results, the agreement becomes slightly worse but remains at the 1 percent level until $k\sim 8$ h/Mpc.

The level of agreement presented in Fig.~\ref{fig:best_fit_7param} is comparable to the results from Ref.~\cite{Arico2021SimultaneousBaryons} that are based on an alternative implementation of the baryonification method \cite{arico2020modelling}. Note however, that both these results are based on a 7-parameter model implementation. For practical purposes, it will be important to further reduce the \RefReport{number} 
of free parameters. We will show results for reduced sets of parameters in Sec.~\ref{sec:reduce_param}.

\subsection{Predicting the power spectrum from halo properties}\label{sec:predpowspec}

\begin{figure}
     \centering
    \begin{subfigure}[b]{0.49\textwidth}
         \centering
         \caption{cosmo-OWLS 8.0}
         \includegraphics[width=\textwidth]{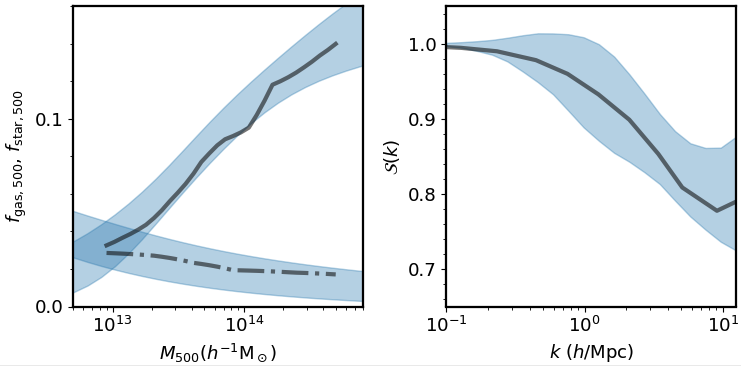}
         \label{fig:cosmoOWLS8p0_fgasfstar_2_ps}
     \end{subfigure}
     \hfill
     \begin{subfigure}[b]{0.49\textwidth}
         \centering
         \caption{cosmo-OWLS 8.5}
         \includegraphics[width=\textwidth]{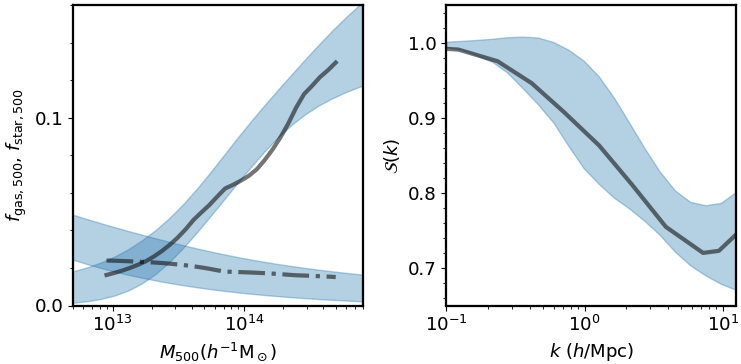}
         \label{fig:cosmoOWLS8p5_fgasfstar_2_ps}
     \end{subfigure}
     \hfill
     \begin{subfigure}[b]{0.49\textwidth}
         \centering
        \caption{BAHAMAS}
         \includegraphics[width=\textwidth]{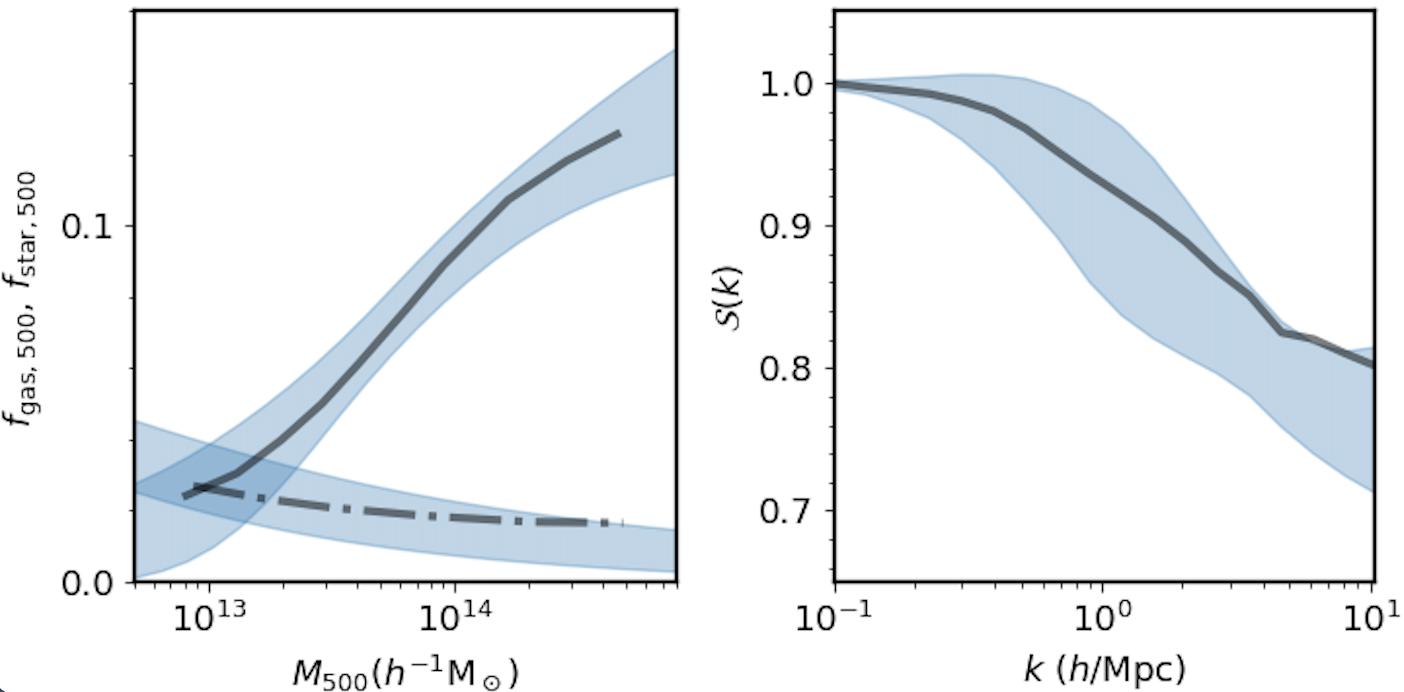}
         \label{fig:BAHAMAS_fgasfstar_2_ps}
     \end{subfigure}
     \hfill
     \begin{subfigure}[b]{0.49\textwidth}
         \centering
         \caption{OWLS}
         \includegraphics[width=\textwidth]{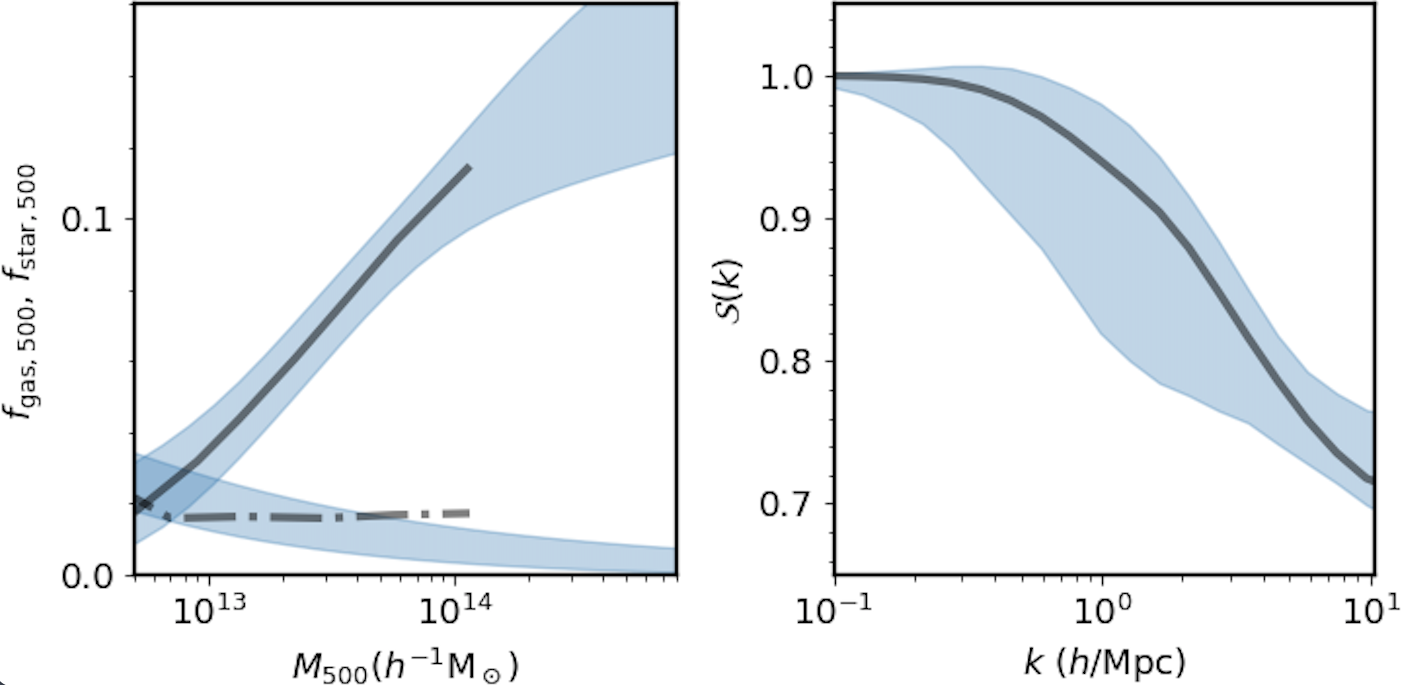}
         \label{fig:OWLS_fgasfstar_2_ps}
     \end{subfigure}
     \hfill
     \begin{subfigure}[b]{0.49\textwidth}
         \centering
         \caption{Illustris-TNG}
         \includegraphics[width=\textwidth]{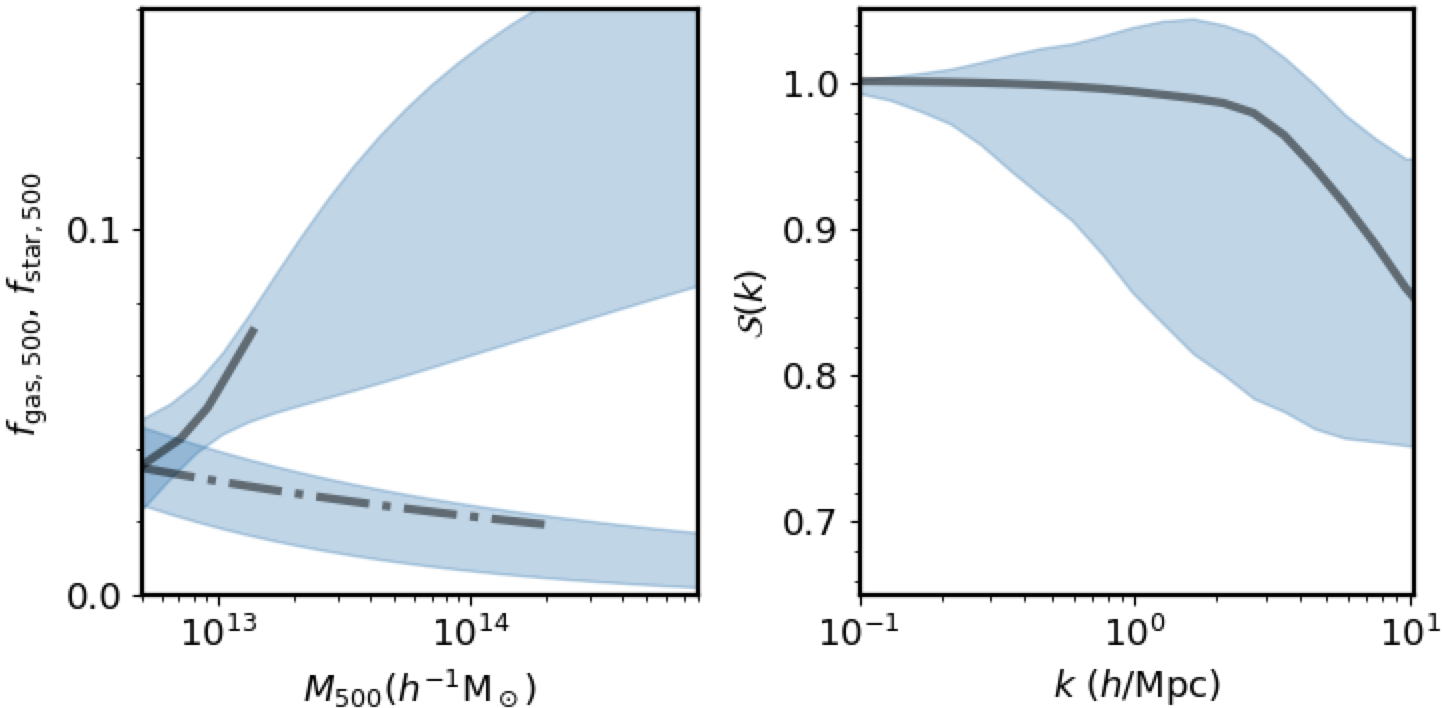}
         \label{fig:IllustrisTNG_fgasfstar_2_ps}
     \end{subfigure}
     \hfill
     \begin{subfigure}[b]{0.49\textwidth}
         \centering
         \caption{Horizon-AGN}
         \includegraphics[width=\textwidth]{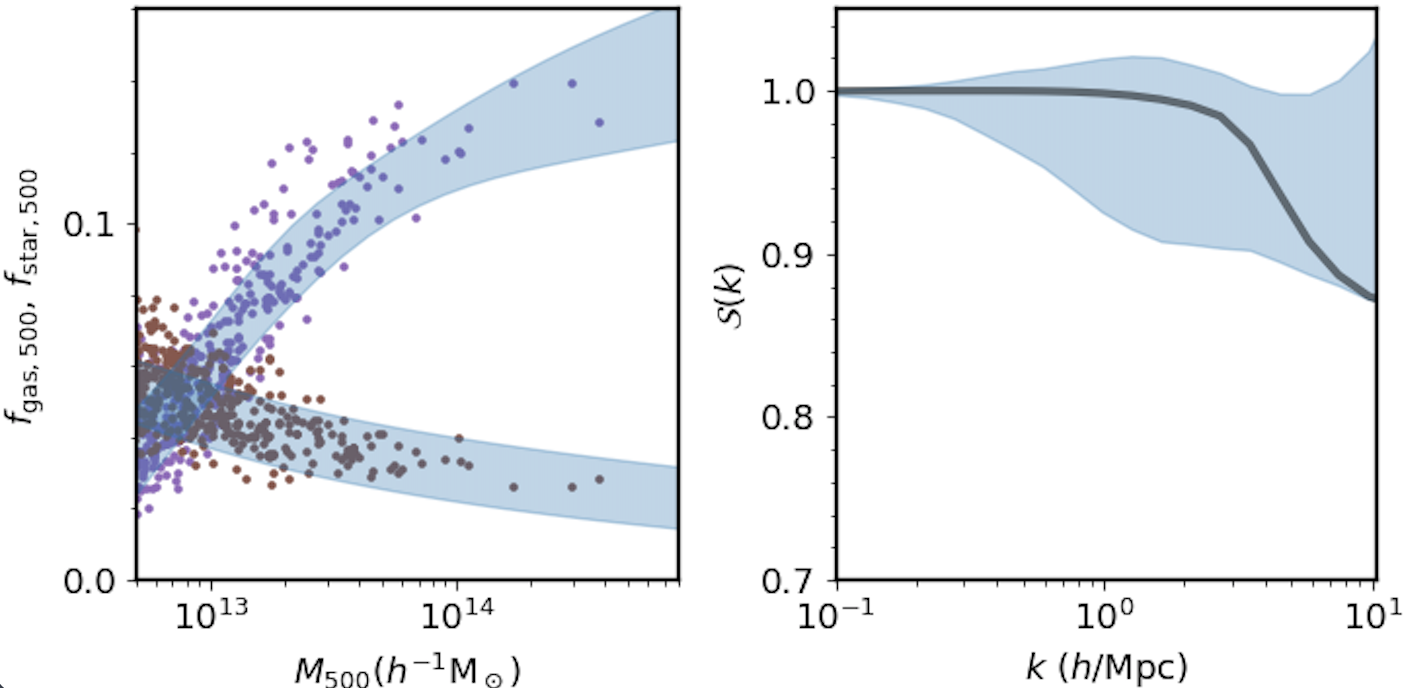}
         \label{fig:HorizonAGN_fgasfstar_2_ps}
     \end{subfigure}
        \caption{Comparison between the baryonification model (blue bands) and the hydro-dynamical simulations {\tt cosmo-OWLS}, {\tt BAHAMAS}, {\tt OWLS}, {\tt Illustris-TNG}, and {\tt Horizon-AGN} (black lines). For each sub-figure, we show both the gas \RefReport{($f_{\rm gas,500}$; solid line)} and stellar \RefReport{($f_{\rm star,500}$; dash-dotted line)} fractions 
        as a function of $M_{500}$ (left-hand panel) as well as the baryonic power suppression ($\mathcal{S}$) as a function of $k$ (right-hand panel). \RefReport{Here} the baryonification model is only fitted to the gas and stellar fractions from simulations. The power suppression is then predicted by the baryonification model (based on the fitted parameter values) and compared to the simulation results. This procedure validates the ability of the baryonification model to recover $\mathcal{S}(k)$ based solely on information of the fractions $f_{\rm gas,500}$ and $f_{\rm star,500}$.}
        \label{fig:fit_hydroSims_fgasfstar_2_ps}
\end{figure}

Another important feature of the baryonification model is its ability to provide a direct link between individual halo profiles and the total matter power spectrum. In this section we will investigate this connection by comparing the baryonifiaction model to the simulations.

The question we want to examine is to what extent the baryonification model is able to reproduce the baryonic power suppression from \RefReport{simulations that are} based solely on the simulated mean gas fractions ($f_\mathrm{gas,500}$) and stellar fractions ($f_\mathrm{star,500}$) within haloes (more precisely within the radius $r_{500}$). We calculate the stellar and gas fractions using the baryonification profiles (see Eq,~\ref{rhogas}) and fit them to the simulated fractions from {\tt OWLS}, {\tt cosmo-OWLS}, {\tt BAHAMAS}, {\tt IllustrisTNG}, and {\tt HorizonAGN}. For the fitting procedure, we consider 20 bins between $10^{13} h^{-1}M_\odot$ and $2\times 10^{14} h^{-1}M_\odot$ and assume a standard deviation of 1 percent 
on both $f_\mathrm{gas,500}$ and $f_\mathrm{star,500}$. We ignore the bins where the data from simulations are not available.
We then run an MCMC to fit all seven parameters simultaneously. From the resulting chain, we take parameter values within the 68 percent credible regions of the posterior distributions to determine the corresponding $\mathcal{S}(k)$ using our emulator.

The gas and stellar fractions fitted to each simulation (left-hand panels) as well as the corresponding prediction for the power spectrum suppression (right-hand panel) are shown in Fig~\ref{fig:fit_hydroSims_fgasfstar_2_ps}. While the black lines represent the results from simulations, the coloured bands indicate the 68-percent contours of the baryonification model. Fig.~\ref{fig:fit_hydroSims_fgasfstar_2_ps} shows that the baryonification model is able to reproduce the shapes and amplitudes of the simulated baryonic power suppression to good accuracy when fitted to the gas and stellar fractions from simulations. This is an important result validating the methodology of the baryonification method. Note that the results also agree with the analysis of Ref.~\cite{Schneider2019QuantifyingCorrelation}, where a similar exercise has been carried out, albeit using a much less general and robust fitting procedure. 

A closer look at Fig.~\ref{fig:fit_hydroSims_fgasfstar_2_ps} reveals that the agreement is best for the {\tt OWLS} and {\tt cosmo-OWLS} simulations. The {\tt BAHAMAS} simulations can also be recovered well, albeit only when correcting for resolution effects at small scales. Indeed, it is shown in Ref.~\cite{vanDaalen2019ExploringSpectra} that the {\tt BAHAMAS} power spectrum is biased due to resolution effects above $k\sim 5$ h/Mpc. We correct for this bias in Fig.~\ref{fig:BAHAMAS_ps_2_fgasfstar} by multiplying the original curve with the resolution-corrected difference provided in appendix A of Ref.~\cite{vanDaalen2019ExploringSpectra}.

\RefReport{For the {\tt Illustris-TNG} simulation shown in Fig.~\ref{fig:IllustrisTNG_fgasfstar_2_ps}, the constraints on the power suppression remain rather poor. This is a direct result of the fact that we only have information about gas fractions from relatively small halo masses (of $M_{500}\lesssim 2\times 10^{13}~h^{-1}M_\odot$). 
Note that the power spectrum \cite{springel2018first} and the stellar fractions \cite{pillepich2018StellarMass} are obtained from the {\tt Illustris-TNG} simulation with box-length of $L=75~h^{-1}$ Mpc. The gas fractions, on the other hand, are obtained from Ref.~\cite{pillepich2018simulating}, which is based on a simulation with the same sub-grid modelling but a smaller box-size of $L=25~h^{-1}$ Mpc. 
}


For the {\tt Horizon-AGN} simulation illustrated in Fig.~\ref{fig:HorizonAGN_fgasfstar_2_ps}, we use the $f_\mathrm{gas,500}$ and $f_\mathrm{star,500}$ values of individual haloes instead of binned quantities (see coloured points in the left-hand panel). The resulting constraints on the power suppression are rather poor \RefReport{as the likelihood is worse} \RefReportTwo{by about 100 times compared to the other simulations}. 
We speculate that this could be a consequence of the large scatter in the individual gas and stellar fractions from the simulation.





\subsection{Predicting halo properties from the power spectrum}
\label{sec:predgasfrac}
In the previous section we have shown that the baryonification model is able to reproduce the baryonic power suppression based on the stellar and gas properties of haloes. We now want to investigate the inverse procedure, i.e. to test if the model can reproduce the gas and stellar fractions from simulations when fitted to the baryonic power suppression. For the fitting procedure, we use a setup similar to the one above, allowing for a 1 percent error on the wave-modes between 0.1 and 12.5 h/Mpc.

\begin{figure}
     \centering
    \begin{subfigure}[b]{0.49\textwidth}
         \centering
         \caption{cosmo-OWLS 8.0}
         \includegraphics[width=\textwidth]{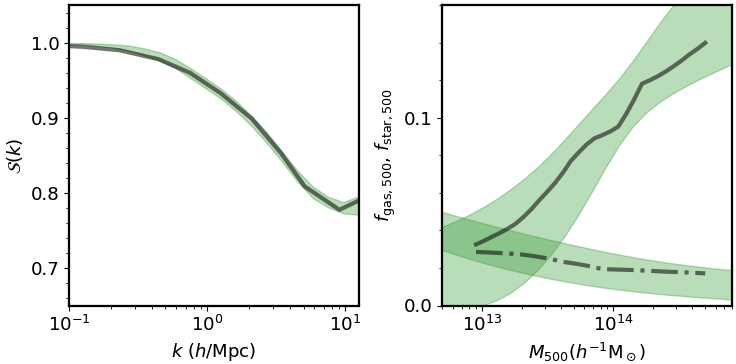}
         \label{fig:cosmoOWLS8p0_ps_2_fgasfstar}
     \end{subfigure}
     \hfill
     \begin{subfigure}[b]{0.49\textwidth}
         \centering
         \caption{cosmo-OWLS 8.5}
         \includegraphics[width=\textwidth]{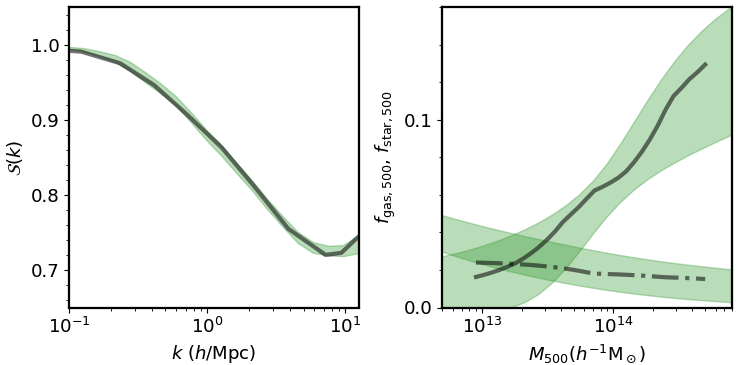}
         \label{fig:cosmoOWLS8p5_ps_2_fgasfstar}
     \end{subfigure}
     \hfill
     \begin{subfigure}[b]{0.49\textwidth}
         \centering
        \caption{BAHAMAS}
         \includegraphics[width=\textwidth]{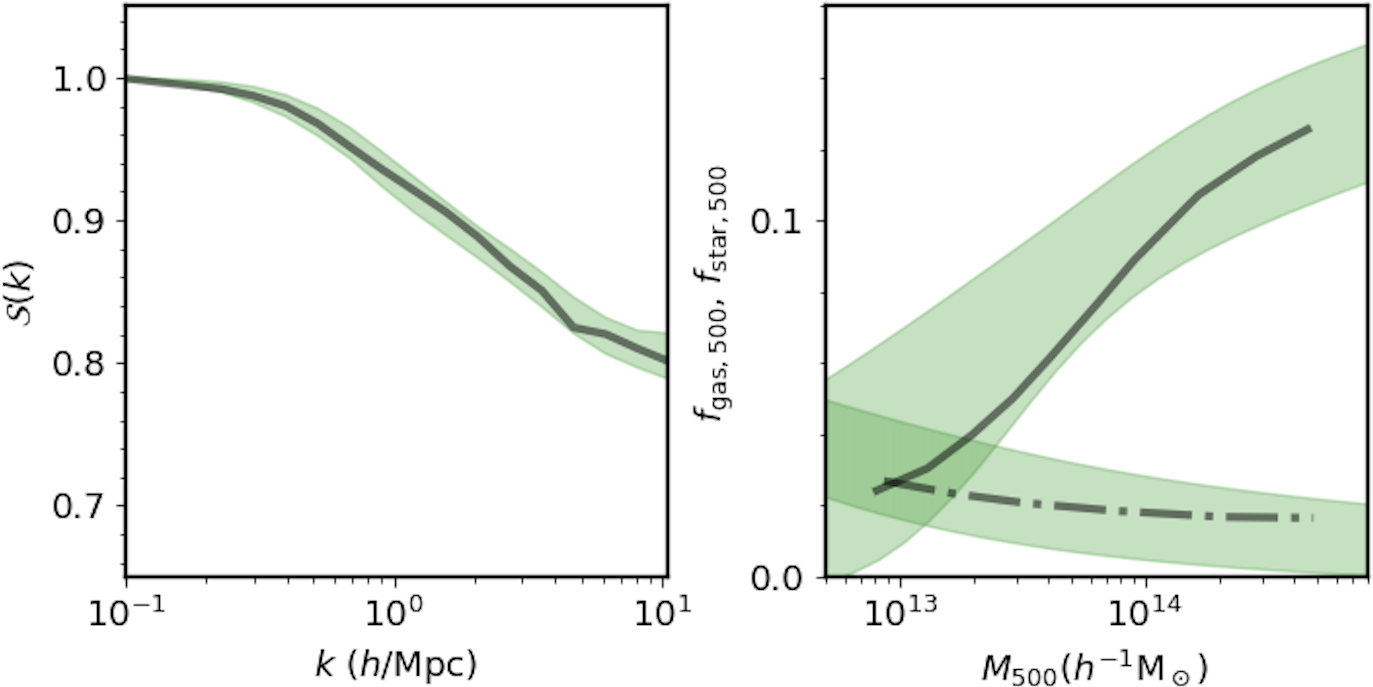}
         \label{fig:BAHAMAS_ps_2_fgasfstar}
     \end{subfigure}
     \hfill
     \begin{subfigure}[b]{0.49\textwidth}
         \centering
         \caption{OWLS}
         \includegraphics[width=\textwidth]{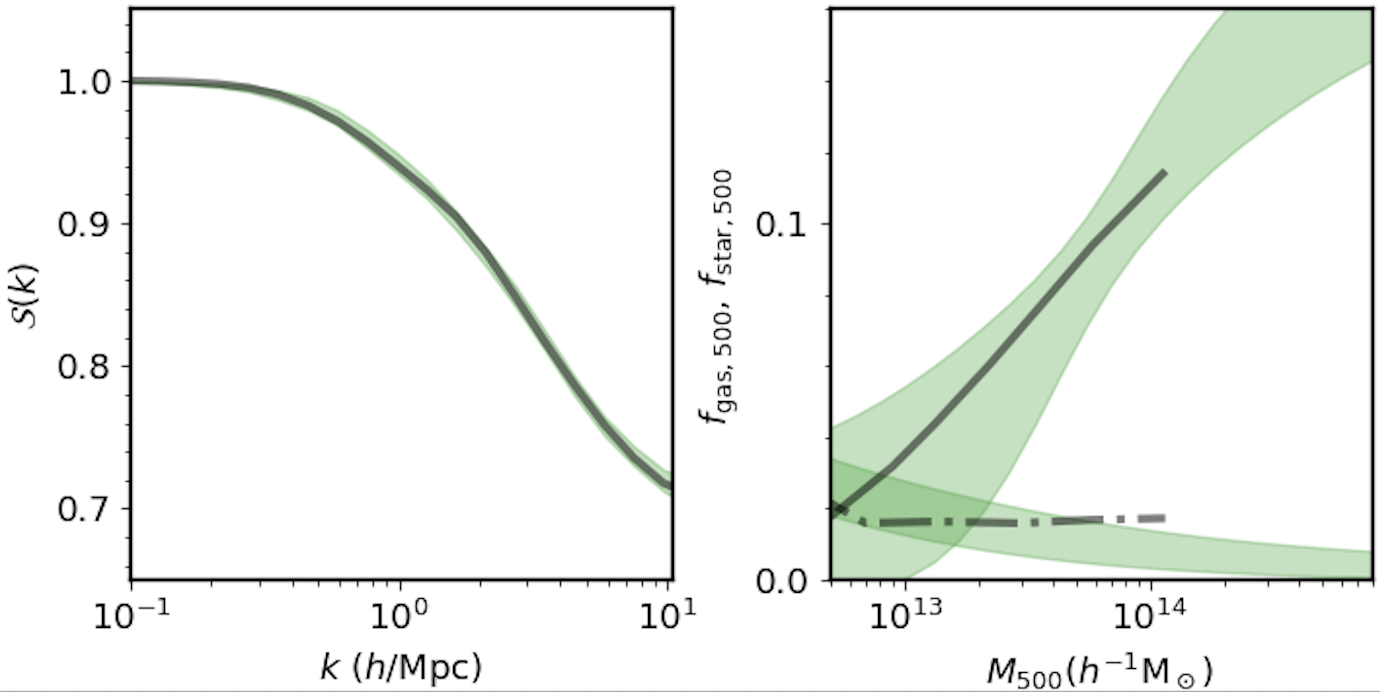}
         \label{fig:OWLS_ps_2_fgasfstar}
     \end{subfigure}
     \hfill
     \begin{subfigure}[b]{0.49\textwidth}
         \centering
         \caption{Illustris-TNG}
         \includegraphics[width=\textwidth]{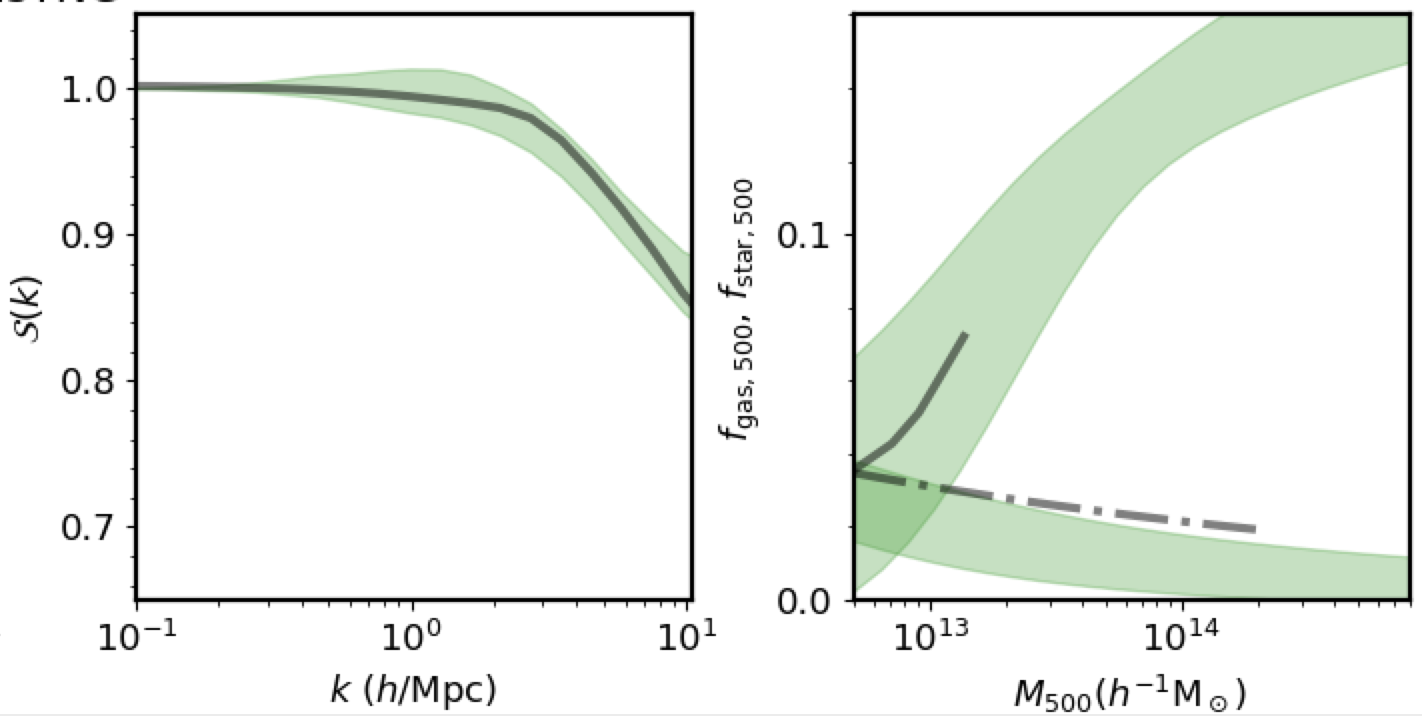}
         \label{fig:IllustrisTNG_ps_2_fgasfstar}
     \end{subfigure}
     \hfill
     \begin{subfigure}[b]{0.49\textwidth}
         \centering
         \caption{Horizon-AGN}
         \includegraphics[width=\textwidth]{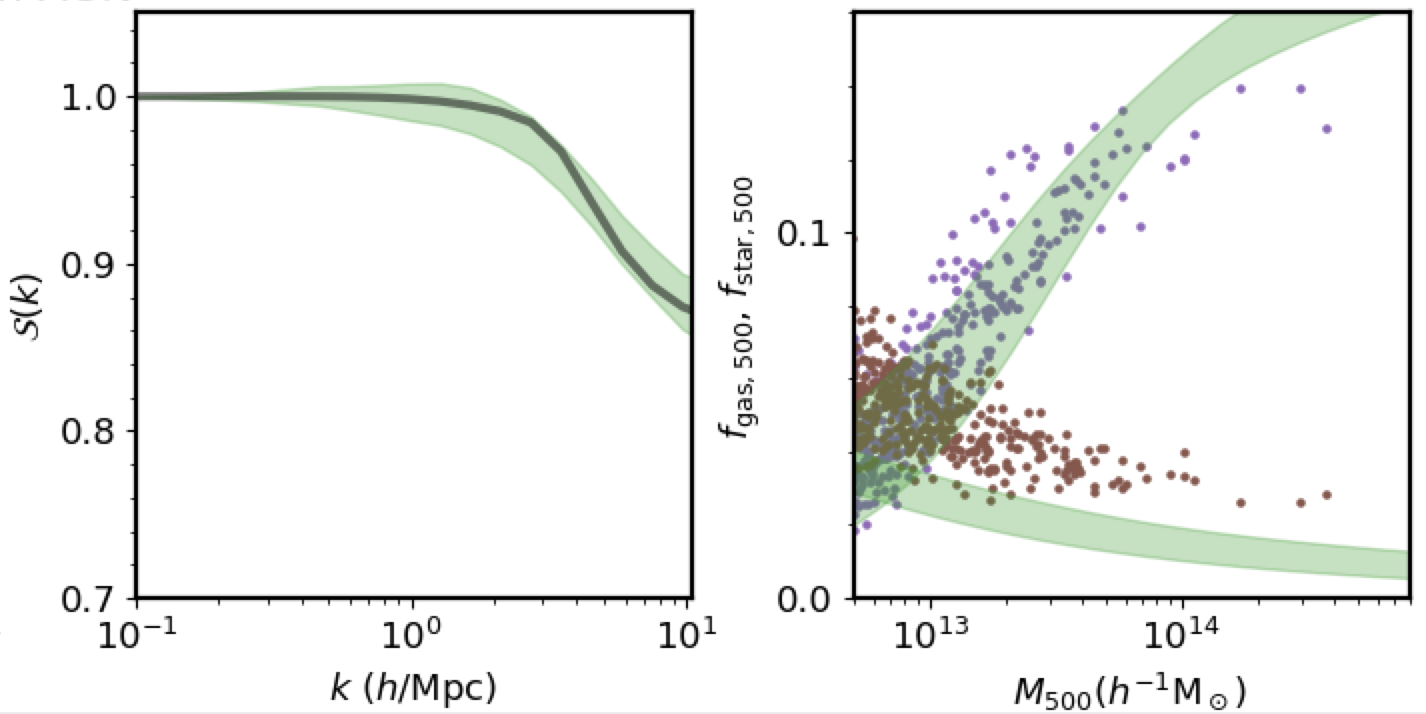}
         \label{fig:HorizonAGN_ps_2_fgasfstar}
     \end{subfigure}
        \caption{Comparison between the baryonification model (green bands) and the hydro-dynamical simulations {\tt cosmo-OWLS}, {\tt BAHAMAS}, {\tt OWLS}, {\tt Illustris-TNG}, and {\tt Horizon-AGN} (black lines). In contrast to Fig.~\ref{fig:fit_hydroSims_fgasfstar_2_ps}, we now fit the baryonification model to the power suppression $\mathcal{S}$ of each simulation (left-hand panel of each sub-figure) with the goal to recover the gas \RefReport{($f_{\rm gas,500}$; solid line)} and stellar \RefReport{($f_{\rm star,500}$; dash-dotted line)} fractions as a function of $M_{500}$ (right-hand panel of each sub-figure). The figure confirms that the baryonification method is able to recover simulated halo properties based on information from the power spectrum.}
        \label{fig:fit_hydroSims_ps_2_fgasfstar}
\end{figure}

The fits of the power suppression ($\mathcal{S}$) and the resulting predictions for the gas ($f_\mathrm{gas,500}$) and stellar fractions ($f_\mathrm{star,500}$) are shown in Fig.~\ref{fig:fit_hydroSims_ps_2_fgasfstar} as green bands. The width of the bands are obtained by drawing baryonification parameters from 68 percent credible regions. The corresponding measures from hydro-dynamical simulations are provided as black lines. As before, each simulation is accounted for in a separate sub-figure (a-e).

The best results are again obtained for {\tt cosmo-OWLS} (Fig.~\ref{fig:cosmoOWLS8p0_ps_2_fgasfstar}, \ref{fig:cosmoOWLS8p5_ps_2_fgasfstar}), {\tt BAHAMAS} (Fig.~\ref{fig:BAHAMAS_ps_2_fgasfstar}), and {\tt OWLS} (Fig.~\ref{fig:OWLS_ps_2_fgasfstar}), where the baryonification model is able to recover the simulated gas and stellar fractions over the full mass range. Note that for the {\tt BAHAMAS} simulation, we fit to the resolution-corrected suppression signal of the power spectrum (see Sec.~\ref{sec:predpowspec} and Appendix A of Ref.~\cite{vanDaalen2019ExploringSpectra} for more details). Ignoring this correction would lead to a visible bias between the gas and stellar fractions from the simulation and the baryonification model.

For the {\tt Illustris-TNG} and the {\tt Horizon-AGN} simulations shown in Fig.~\ref{fig:IllustrisTNG_ps_2_fgasfstar} and \ref{fig:HorizonAGN_ps_2_fgasfstar}, we find good agreement for the gas fraction predicted by the baryonificatin model. The stellar fraction, on the other hand, is predicted to be somewhat smaller than the simulation results. At least for {\tt Illustris-TNG}, this could be due to the fact that the stellar fraction comes from a lower-resolution simulation (see discussion in Sec.~\ref{sec:predpowspec}). 


\subsection{Predicting the cluster baryon fraction from the power spectrum}
The comparison tests carried out in the two previous sections are limited to the six simulations with published data for $f_{\rm gas,500}$ and $f_{\rm star,500}$ as function of $M_{500}$ (presented in Table \ref{table:hydro_sims}).  However, if we restrict ourselves to the mean baryon fraction of clusters, then there are more simulations to investigate. 
Indeed, Ref.~\cite{vanDaalen2019ExploringSpectra} published the baryonic fractions at $M_{500}=10^{14}$ M$_{\odot}$, $f_b(\sim 10^{14} M_{\odot})$, for the six simulations introduced above plus the additional simulations {\tt BAHAMAS low-AGN}, {\tt BAHAMAS high-AGN}, and {\tt Illustris}\footnote{The {\tt BAHAMAS low-AGN} and {\tt BAHAMAS high-AGN} simulations assume 0.2 dex lower and higher AGN heating temperatures compared to the original {\tt BAHAMAS}. For details about the {\tt Illustris} simulations, see Refs.~\cite{vogelsberger2014properties,vogelsberger2014introducing,genel2014introducing})}. We will now use this extended data-set to check how well the baryonification model is able to predict the value of $f_b(\sim 10^{14} M_{\odot})$ when fitted to the baryonic power suppression ($\mathcal{S}$).

\begin{figure}[t] 
 \centering
  \includegraphics[width=0.65\textwidth]{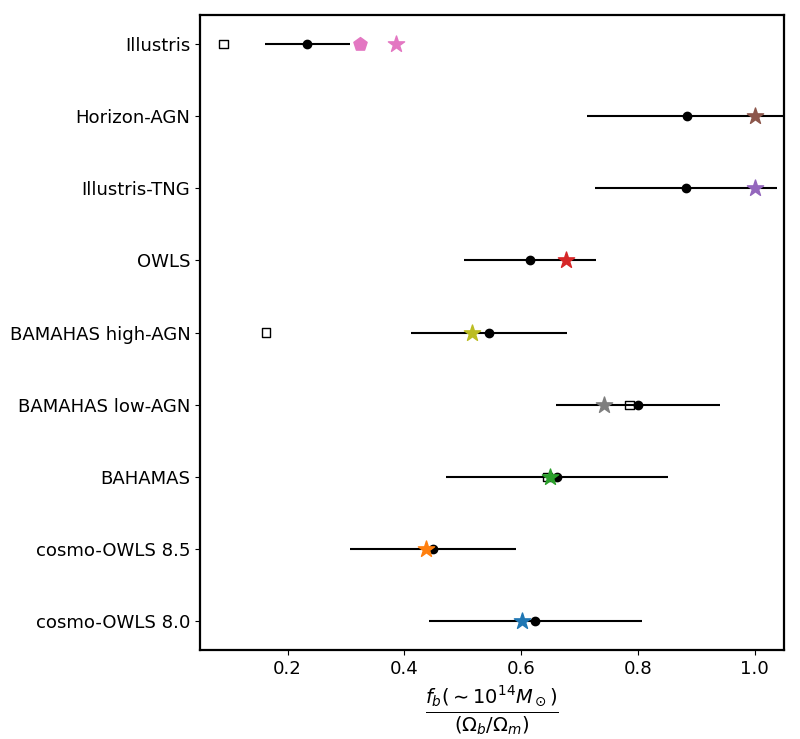}
 \caption{Mean cluster baryon fractions for haloes with $M_{500}\sim 10^{14} M_\odot$ in various hydrodynamical simulations (coloured symbols). While the stars represent the values published in Ref.~\cite{vanDaalen2019ExploringSpectra}, the pink pentagon is obtained from Ref.~\cite{haider2016large}. The black data points with error bars show predictions from our baryonification model obtained by fitting the model to the baryonic power suppression from each simulation. The size of the error-bars corresponds to the uncertainty-range of the fitting procedure (see text for more details). Note that all simulation results are well recovered by the baryonification model, with only the {\tt Illustris} results lying outside the 1$\sigma$ uncertainty range. The empty squares give the values estimated in Ref.~\cite{Arico2021SimultaneousBaryons} based on a different implementation of the baryonification method.
 }
\label{fig:estimated_baryon_fractions}
\end{figure}

In Fig.~\ref{fig:estimated_baryon_fractions}, we show the mean baryonic fractions $f_b(\sim 10^{14} M_{\odot})$ provided by Ref.~\cite{vanDaalen2019ExploringSpectra} (coloured stars) and we compare them to the results from the baryonification model (black data points). The latter \RefReport{results} are obtained by fitting the baryonification model to the power suppression ($\mathcal{S}$) of the corresponding simulations, before predicting $f_b(\sim 10^{14} M_{\odot})$ based on the obtained best-fitting parameters. The shown error-bars correspond to the 68 percent confidence level based on the fits to $\mathcal{S}$ assuming 1 percent errors on each $k$-mode in the range $k\sim0.1-12.5$ h/Mpc (same as in Sec.~\ref{sec:predgasfrac}).

Fig.~\ref{fig:estimated_baryon_fractions} highlights the very good agreement between the cluster baryon fractions predicted by the simulations and obtained with the baryonification model. Nearly all of the simulation results comfortably lie within the 68 percent uncertainty regime. Only {\tt Illustris} has a cluster baryon fraction that is about $2\sigma$ higher than the prediction by the baryonification model. Note, however, that the difference is reduced to a little more than 1$\sigma$ when \RefReport{compared} to the result from Ref.~\cite{haider2016large} (pink pentagon) instead of Ref.~\cite{vanDaalen2019ExploringSpectra} (pink star). In Ref.~\cite{haider2016large} it is furthermore shown that, due to the small simulation box ($75~h^{-1}$ Mpc) of the {\tt Illustris} simulation, the recovered function $f_b(M_{500})$ becomes very noisy around the relevant scale of $10^{14}$ $M_{\odot}$. This points towards the underlying reason for the difference between \cite{vanDaalen2019ExploringSpectra} and \cite{haider2016large}, letting us conclude that the cluster baryon fraction from {\tt Illustris} remains somewhat uncertain.

Note that for the {\tt Illustris} and the three {\tt BAHAMAS} simulations, we have added the published results from Ref.~\cite{Arico2021SimultaneousBaryons} (empty squares) that are based on a different implementation of the baryonification method. While we obtain very similar predictions for {\tt BAHAMAS} and {\tt BAHAMAS low-AGN}, there are notable differences for the {\tt BAHAMAS high-AGN} and {\tt Illustris} simulations, with the results from Ref. \cite{Arico2021SimultaneousBaryons} significantly deviating from the simulations and from our results.

\section{Constraints on baryonic effects with X-ray observations}
\label{sec:constraints_obs}

In the previous section we have validated the baryonifcation method as a tool to recover the baryonic power suppression ($\mathcal{S}$) based on information about the properties of haloes around galaxy groups and clusters. We will now use this connection to predict $\mathcal{S}$ based on observed gas and stellar fractions from X-ray and optical observations. In Sec.~\ref{sec:constraints_on_bfc_param} we first quantify the constraints on the baryonic parameters, before predicting the shape of the power spectrum in Sec.~\ref{sec:constraints_on_Sk}. Throughout this analysis, we will fix all cosmological parameters to the values from \texttt{Planck} \cite{planck2016CosmologicalParameters}.

 \begin{figure}[t] 
 \centering
  \includegraphics[width=0.75\textwidth]{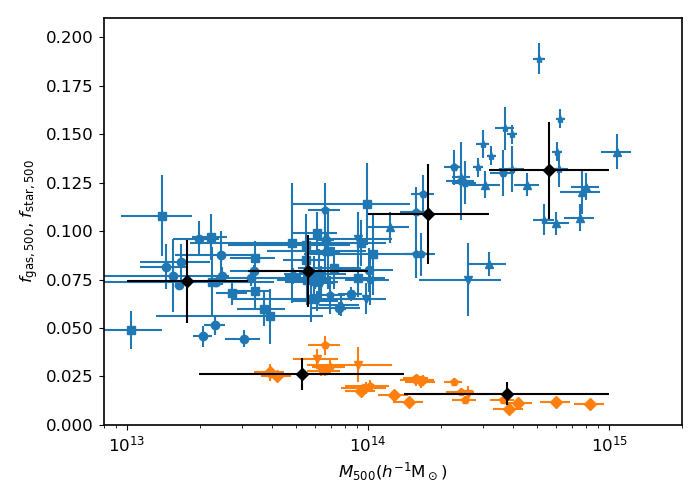}
 \caption{Gas and stellar fractions determined from X-ray and optical observations of galaxy groups and clusters \cite{vikhlinin2006chandra,gonzalez2013galaxy,sanderson2013baryon,lovisari2015scaling,Kravtsov2018StellarHalos}. We have binned the gas and stellar fractions into four and two bins, respectively. The error bars on the binned data includes both the observational errors on the points as well as their scatter.
 }
 	\label{fig:xray_fgas500_fstar500_binned}
 \end{figure}

\subsection{Constraints on baryonic parameters}
\label{sec:constraints_on_bfc_param}
The parameters of the baryonification model will be constrained using gas and stellar fractions of individual galaxy groups and clusters. We thereby follow a similar strategy to that in Ref.~\cite{Schneider2019QuantifyingCorrelation}, but we will perform a more systematic multi-parameter analysis using Bayesian inference.

Fig.~\ref{fig:xray_fgas500_fstar500_binned} shows all the observational data in terms of gas (blue points) and stellar fractions (orange points) that \RefReport{are} used in this work. The data is taken from Refs.~\cite{vikhlinin2006chandra,gonzalez2013galaxy,sanderson2013baryon,lovisari2015scaling,Kravtsov2018StellarHalos} and contains estimates for the observational errors. We furthermore expect the scatter of the data points to depend on the stochastic nature of structure formation, where haloes of the same mass may contain different amounts of gas and stars.

For our analysis, it is convenient to bin the gas and stellar fractions of individual clusters into four and two halo mass bins, respectively. These bins are shown as black data points in Fig.~\ref{fig:xray_fgas500_fstar500_binned}. The error on the binned data accounts for both the observational error and scatter of the points inside each bin. In our analysis, we assume the errors on the binned data to be uncorrelated.

In order to obtain gas and stellar fraction with X-ray observations, the total halo mass has to be estimated assuming the gas to be in hydrostatic equilibrium.
This assumption is, however, known to be sub-optimal because of non-thermal pressure contributions from in-falling gas. Various studies have shown that mass estimates based on hydrostatic equilibrium are likely to underestimate the true cluster mass by 5-30 percent \cite[e.g.][]{sereno2015comparing,eckert2019non,ettori2019hydrostatic}. Following the usual practice, we  quantify this uncertainty using a hydrostatic mass bias ($b_\mathrm{hse}$) that connects the estimated to the true mass via the relation
\begin{eqnarray}\label{hydrostaticmass}
 M_\mathrm{500,hse} = (1-b_\mathrm{hse})M_\mathrm{500} \ .
\end{eqnarray}
Here $M_\mathrm{500,hse}$ is the total mass estimated assuming hydrostatic equilibrium and $M_\mathrm{500}$ is the true total mass inside of $r_\mathrm{500}$.

 \begin{figure}[t] 
 \centering
    \includegraphics[width=1.0\textwidth]{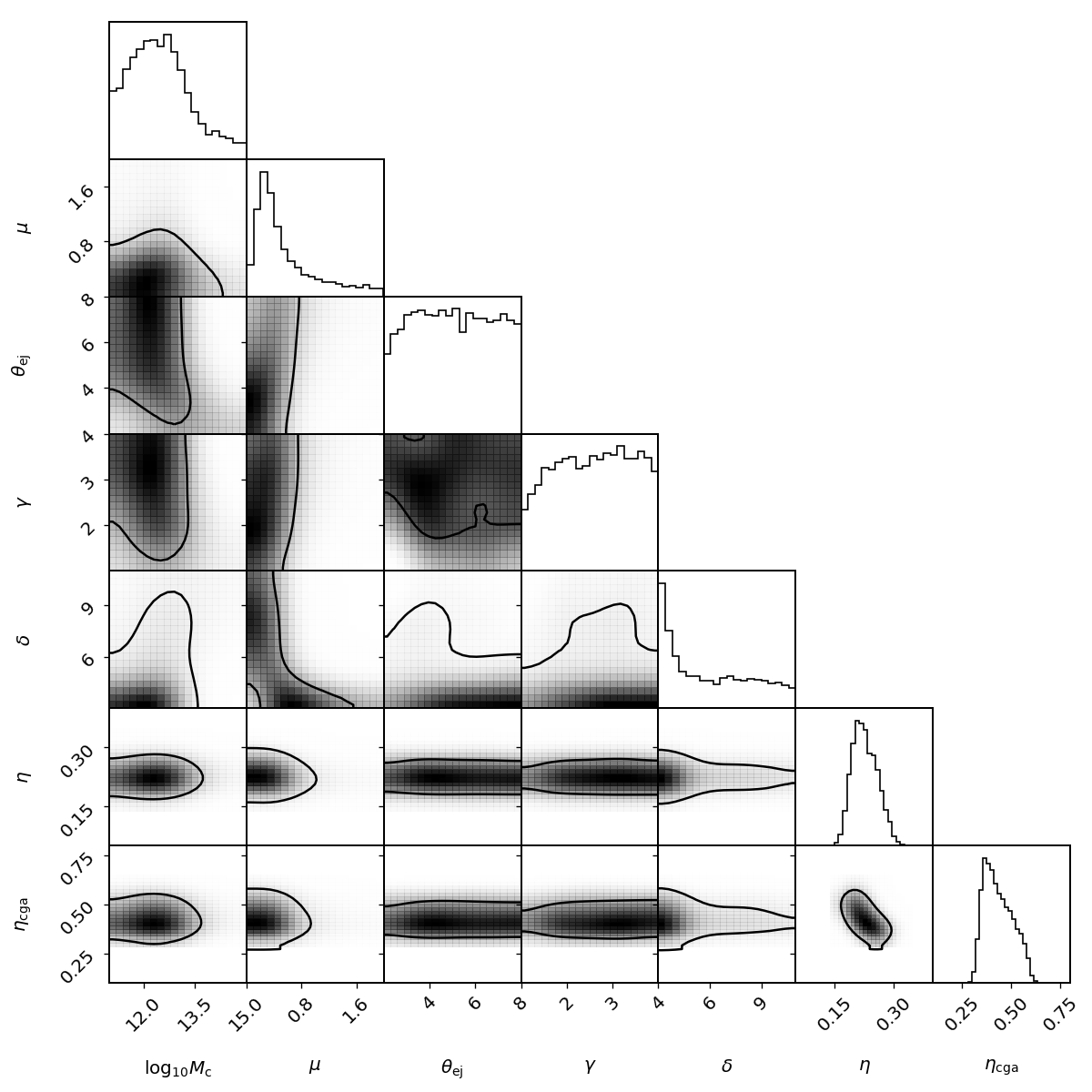}

 \caption{Posterior contours of the seven parameters from the baryonification model fitted to X-ray and optical data of gas and stellar fraction within the halo radius $r_{500}$. Here we have fixed all cosmological parameters to the best-fitting values from {\tt Planck} and we have marginalised over the hydrostatic mass bias parameter ($b_\mathrm{hse}$).}
 	\label{fig:corner_BCMparam_xray}
 \end{figure}

We infer constraints in the baryonic feedback processes by running MCMC using our seven-parameter baryonification model. For all baryonic parameters, we consider flat priors within the range given in Table~\ref{table:range_values}.
For the hydrostatic mass bias ($b_\mathrm{hse}$) we use a Gaussian prior with mean and standard deviation of 0.26 and 0.07, respectively. These values are taken from the analysis of Ref.~\cite{Hurier2017MeasuringData} that is based on CMB lensing and Sunyaev-Zeldovich data. Note that these values are in good agreement with other direct estimates based on X-ray and lensing data \cite[e.g.][]{douspis2018tension}.

In Fig.~\ref{fig:corner_BCMparam_xray}, we plot the resulting posterior distribution from our MCMC run (marginalising over $b_\mathrm{hse}$). The parameter log$_\mathrm{10} M_c$ is constrained to a value around 12.6, which is similar to the standard value used in previous studies \cite{Schneider2019QuantifyingCorrelation,Schneider2019BaryonicMatrix}. For the second parameter $\mu$, the data prefer low values, which is a consequence of the fact that the X-ray gas fraction is only weakly dependent on halo mass. The parameters $\theta_\mathrm{ej}$ and $\gamma$ are not well constrained by the data. The parameter $\delta$, on the other hand, is pushed to rather low values, corresponding to a slowly falling gas profile beyond the virial radius (see Eq.~\ref{rhogas}). This means that large amounts of gas are pushed outside of the halo, providing evidence for strong feedback processes.


\RefReport{Fig.~\ref{fig:corner_BCMparam_xray} shows that the stellar parameters are well constrained by the galaxy cluster observations. Note that the parameter $\eta_\mathrm{cga}=\eta+\eta_\delta$ is a derived quantity. We constrain stellar parameters $\eta$ and $\eta_\mathrm{cga}$ to be $0.23\pm 0.03$ and $0.42\pm 0.07$, respectively.}

 \begin{figure}[t] 
 \centering
   \includegraphics[width=1.0\textwidth]{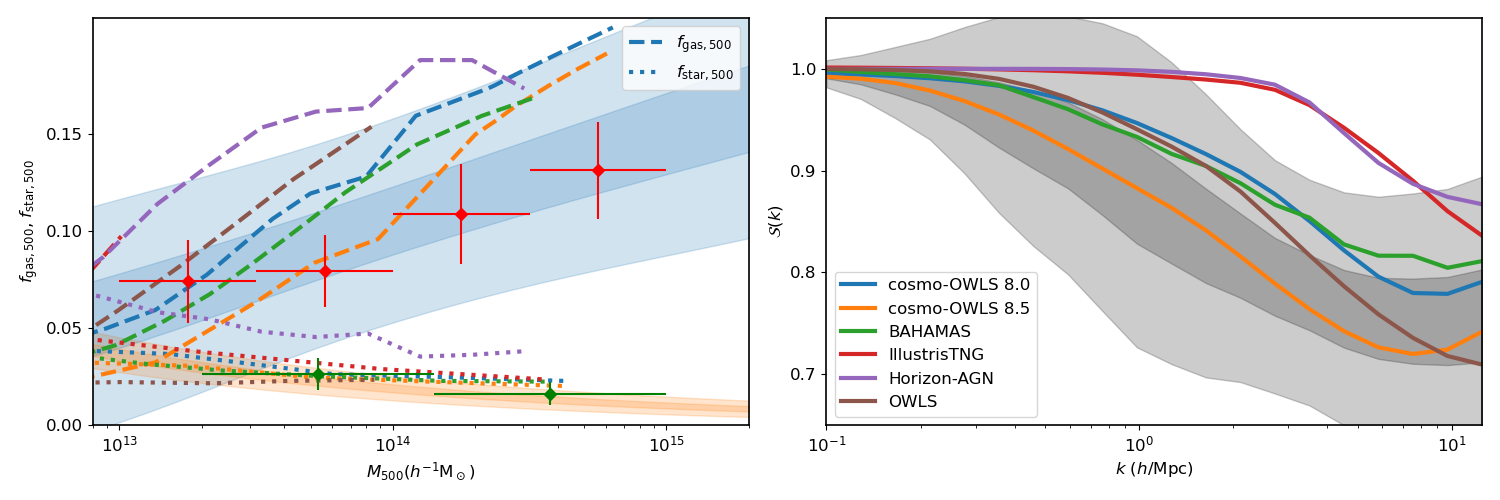}
 \caption{Constraints on $f_\mathrm{gas}$, $f_\mathrm{star}$, and $\mathcal{S}(k)$ based on the X-ray observations. The left-hand panel gives the binned X-ray data for the gas (red points) and stellar (green points) fractions at different cluster masses. The dark and light shaded regions show the 68 and \RefReport{99} percent credible regions determined from our MCMC run. 
 \RefReport{The dashed and dotted lines represent the gas and stellar fractions from different hydro-dynamical simulations. Note that the simulated values of $M_{500}$ are corrected according to Eq.~(\ref{hydrostaticmass}), assuming $b_\mathrm{hse}=0.26$ to make them directly comparable to the observations.}
 The right-hand panel shows the 68 and \RefReport{99} percent credible regions of our MCMC constraints on $\mathcal{S}(k)$ as dark and light shaded regions.
 }
 	\label{fig:fitted_fgas_fstar_Sk_BCMparam_xray}
 \end{figure}

\subsection{Constraints on the power spectrum}
\label{sec:constraints_on_Sk}
Using the constraints on the baryonification parameters (shown in Fig.~\ref{fig:corner_BCMparam_xray}), we can now investigate the expected shape of the baryonic power suppression ($\mathcal{S}$). The result can then be compared to the predictions from hydro-dynamical simulations, providing an indirect test for their validity regarding cosmology.

In the left-hand panel of Figure~\ref{fig:fitted_fgas_fstar_Sk_BCMparam_xray}, we show the $f_\mathrm{gas,500}$ and $f_\mathrm{star,500}$ as blue and orange bands. They include the 68 percent (dark band) and \RefReport{99} percent (light band) credible regions obtained via the posterior distribution of the baryonification parameters (see Fig.~\ref{fig:corner_BCMparam_xray}). The plot confirms that the binned X-ray and optical data used for the fitting procedure (shown as red and green symbols) is within the credible region \RefReport{obtained} by the MCMC run. \RefReport{For comparison, we have also plotted the gas (dashed-lines) and stellar (dotted-lines) fractions from the hydro-dynamical simulations. These fractions do not correspond to the direct simulation outputs but are instead constructed assuming Eq.~(\ref{hydrostaticmass}), where $M_{500}$ is taken from the simulations and where we assume $b_\mathrm{hse}=0.26$ for the hydrostatic mass bias. This allows us to directly compare the simulation outputs to both the observations and the results from the emulator.}

\RefReport{
We note that the simulated fractions tend to lie above the binned X-ray observations, especially for masses above $M_{500}\sim 2\times 10^{13}$ M$_{\odot}$/h. The mismatch is not only visible for {\tt Illustris-TNG}, {\tt Horizon-AGN}, and {\tt OWLS}, but also for the simulations {\tt cosmo-OWLS 8.0}, {\tt cosmo-OWLS 8.5}, and {\tt BAHAMAS}, which have been originally calibrated to X-ray observations. This is a direct consequence of the hydrostatic mass bias from Ref.~\cite{Hurier2017MeasuringData}, that is considerably larger than the one predicted by most hydrodynamical simulations \cite[see e.g.][]{salvati2018constraints}.} 



In the right-hand panel of Figure~\ref{fig:fitted_fgas_fstar_Sk_BCMparam_xray}, we show the predicted baryonic power suppression ($\mathcal{S}$) obtained by applying the baryonification model to the X-ray and optical data from individual galaxy groups and clusters. The dark and light bands again correspond to the 68 and \RefReport{99} percent credible regions. At the 68 percent confidence level, the baryonic power suppression $\mathcal{S}$ is predicted to exceed the percent level between $k\sim 0.1$ and $0.4$ h/Mpc growing to a maximum 21-28 percent at around $k\sim 7$ h/Mpc. At the \RefReport{99} percent level, suppression exceeds one percent somewhere between $k\sim0.1$ h/Mpc and $1.3$ h/Mpc reaching a maximum of 14-38 percent beyond $k\sim 7$ h/Mpc. 

The right-hand panel of Fig.~\ref{fig:fitted_fgas_fstar_Sk_BCMparam_xray} also shows the predicted baryonic power suppression signal from various hydro-dynamical simulations. We find that \RefReport{only {\tt cosmo-OWLS 8.5} is fully consistent with our data,  while {\tt cosmo-OWLS 8.0}, {\tt BAHAMAS}, and {\tt OWLS} show a mild disagreement and lie between the 68 percent and 99 percent credible region.} 
\RefReport{The {\tt Illustris-TNG} and {\tt Horizon-AGN} simulations, on the other hand, are inconsistent with the X-ray observation informed predictions from the baryonification model. They show a suppression signal that is shifted to higher $k$-values, which is outside the 99 percent confidence range suggested by X-ray observations.}

\section{Reducing the number of baryonic parameters}
\label{sec:reduce_param}

\begin{figure}
     \centering
     \begin{subfigure}[b]{0.9\textwidth}
         \centering
         \caption{Five parameters}
         \includegraphics[width=\textwidth]{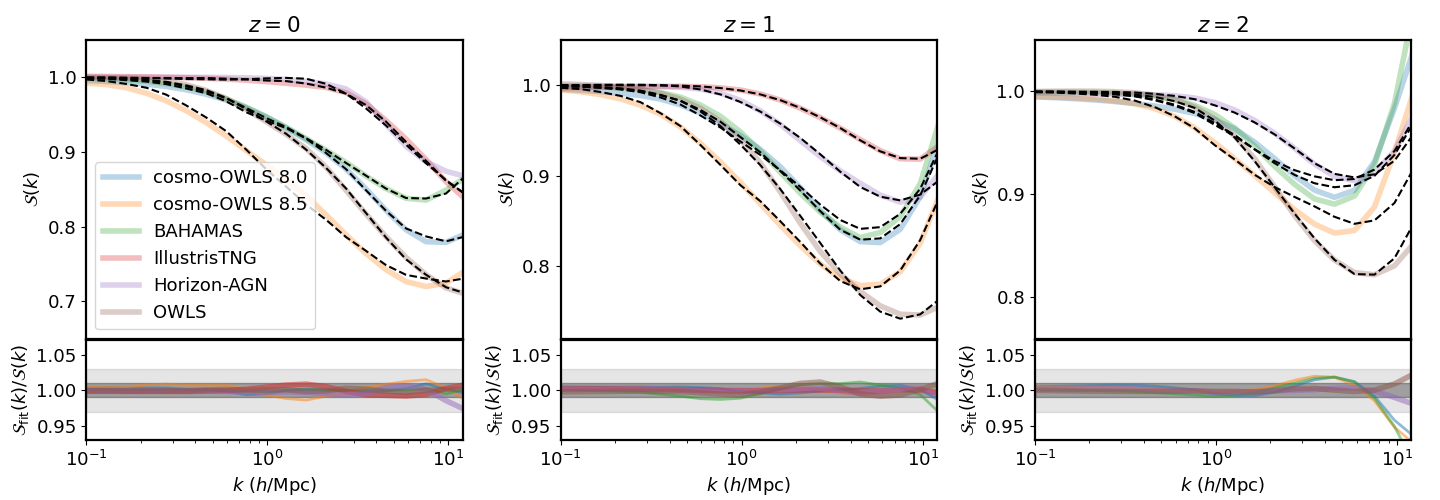}
         \label{fig:best_fit_5param}
     \end{subfigure}
     \hfill
     \begin{subfigure}[b]{0.9\textwidth}
         \centering
        \caption{Three parameters}
         \includegraphics[width=\textwidth]{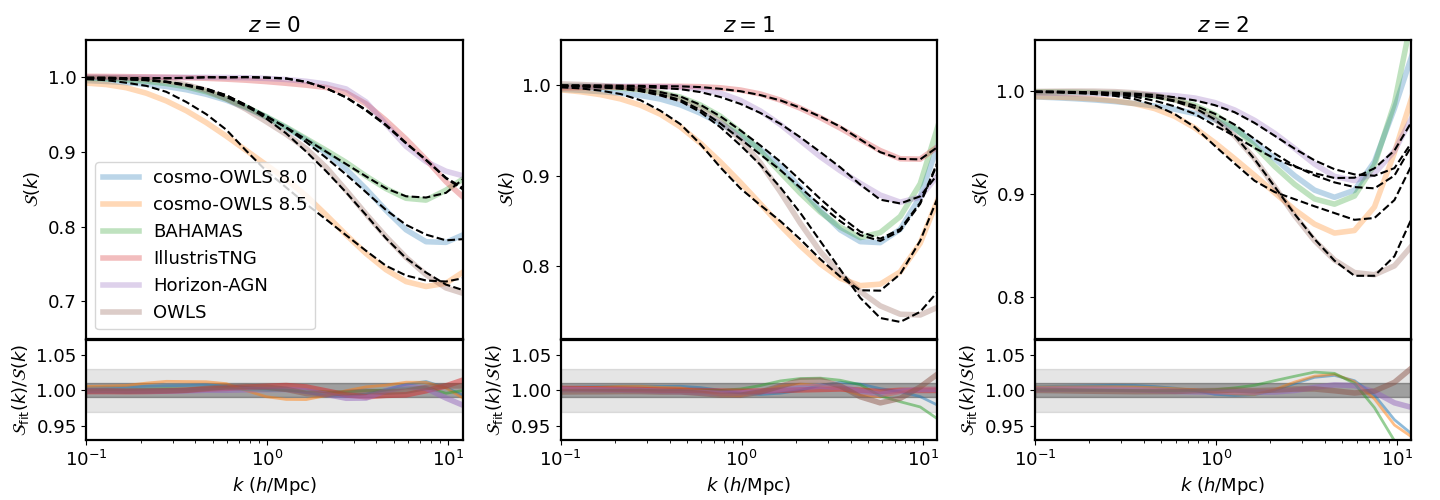}
         \label{fig:best_fit_3param}
     \end{subfigure}
     \hfill
     \begin{subfigure}[b]{0.9\textwidth}
         \centering
         \caption{Single parameter}
         \includegraphics[width=\textwidth]{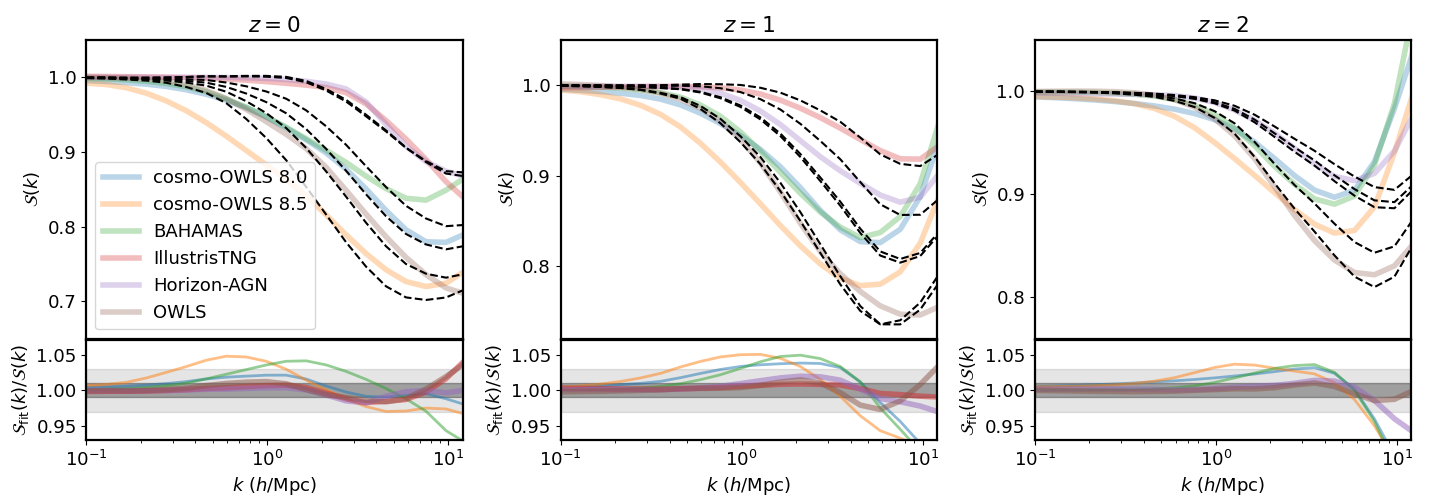}
         \label{fig:best_fit_1param}
     \end{subfigure}
        \caption{Fits to the simulated baryonic power suppression signal ($\mathcal{S}$) for a reduced number of five (a), three (b), and one (c) parameter. From left to right we show results from redshift 0, 1 ,and 2, which are fitted independently. Coloured lines correspond to simulation results, dashed black lines to the emulator. For the one parameter model, we only fit up to $k=3$ h/Mpc, beyond this range the emulator results are shown as dotted lines. For the same plot with the original 7 free parameters, see Fig.~\ref{fig:best_fit_7param}, for details about the parameter reduction procedure, see text and Appendix \ref{sec:model_selection_method}.}
        \label{fig:best_fit_reduce_param}
\end{figure}


Until now we allowed all the seven parameters in our baryonification model to vary. An efficient analysis of cosmological survey data requires, however, a minimum amount of baryonic nuisance parameters. The objective of this section is therefore to find a model with the smallest number of free parameters that still provides a decent fit to the \RefReport{power spectrum suppression} 
from hydro-dynamical simulations. 

In \RefReport{the} first step, we reduce the number of parameters by two, ending up with a five-parameter model. We determine which parameters to fix, by applying a model selection procedure based on the Akaike information criterion (AIC) \cite{akaike1974new,burnham2002practical} and the Bayesian information criterion (BIC) \cite{schwarz1978estimating,burnham2002practical}. A detailed description of the procedure is provided in Appendix~\ref{sec:model_selection_method}.

Our analysis guides us to fix $\delta$ and $\eta$ to 7 and 0.2, respectively. With this five-parameter model, we fit the baryonic power suppression ($\mathcal{S}$) to the predicted signal from the hydro-dynamical simulations introduced in Sec.~\ref{sec:hydro_sims}. The resulting fits (black dashed lines) and simulation results (coloured lines) are shown in Figure~\ref{fig:best_fit_5param}. We find that the five-parameter model provides fits that are nearly as accurate as the original seven-parameter model (see Fig \ref{fig:best_fit_7param}). The agreement is about one percent or better over all scales, except for the $z=2$ case, where we get larger than percent deviations at $k\gtrsim3$ h/Mpc.

As a further step, we reduce the number of parameters from five to three. We do this by fixing $\gamma$ and $\mu$ to 2.5 and 1.0, respectively. The resulting fits for the three-parameter model are shown in Figure~\ref{fig:best_fit_3param}. For redshift 0 and 1, the fits agree to better than 1.5 percent up to $k\sim10$ h/Mpc. For the $z=2$ case, the agreement is better than 3 percent for $k\lesssim 7$ h/Mpc.

Finally, we further reduce the number of free parameters to 1 by fixing $\theta_\mathrm{ej}$ and $\eta_\delta$ to 3.5 and 0.2, respectively. The resulting fits from this 1-parameter model are shown in Figure~\ref{fig:best_fit_1param}. They are significantly less accurate than the ones from the 3-, 5-, and 7-parameter models, showing discrepancies of order 5 percent below $k\sim 8$ h/Mpc (degrading further at higher $k$-modes). We therefore conclude that a model with a single free parameter is insufficient to accurately describe the simulation results for scales up to $k\sim 10$ h/Mpc.
Note, however, that a single parameter model remains a valid option for large scales below $k=2$ h/Mpc. In Appendix~\ref{sec:single_BCM}, we show that for a limited fitting range, 3-percent agreement up to $k\sim2$ h/Mpc can be obtained with a one-parameter model.

\RefReport{We should note that we have reduced the number of parameters based on power spectra from a limited sample of hydro-dynamical simulations. Adding other simulations could alter the conclusions discussed above. The same is true when focusing on quantities unrelated to the power spectrum. Reproducing gas profiles around galaxy clusters, for example, might warrant the use of all seven model parameters \cite[see e.g.][]{Schneider2019QuantifyingCorrelation,schneider2021constraining}.}

\section{Redshift dependence of baryonic effects}
\label{sec:zdep_params}

So far we have ignored the potential redshift dependence of our model parameters. Instead, all models have been fitted to the simulation results at each redshift independently. However, a cohesive framework, where observations at multiple redshifts can be studied simultaneously, may require the baryonification parameters to become functions of redshift \cite[see e.g. Ref.][]{Mead2021Hmcode-2020Feedback}.

In order to study potential redshift dependencies, we will consider the three-parameter model introduced above, as it yields almost same level of accuracy compared to the original seven-parameter model in reproducing the baryonic power suppression signal. In terms of simulations, we focus on {\tt BAMAHAS}, {\tt cosmo-OWLS}, and {\tt OWLS}, which all provide published power spectra at redshifts 0.0, 0.5, 1.0, 1.5, and 2.0.

 \begin{figure}[t] 
 \centering
  \includegraphics[width=1.0\textwidth]{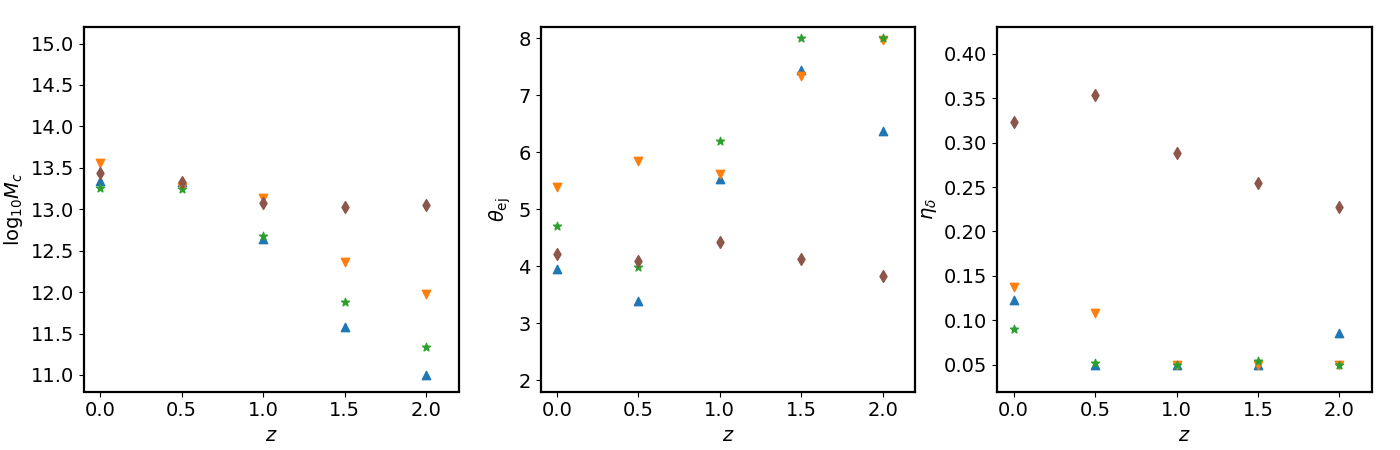}
 \caption{Redshift dependence of the free model parameters (assuming the 3-parameter model), when fitted to each redshift bin individually. Different colours correspond to different fits to the simulations {\tt cosmo-OWLS 8.0} (blue upward-triangles), {\tt cosmo-OWLS 8.5} (orange downward-triangles), {\tt BAMAHAS} (green stars), and {\tt OWLS} (brown diamonds).}
 	\label{fig:3param_vs_z}
 \end{figure}
 
In Fig.~\ref{fig:3param_vs_z}, we show the parameter values (for the three model parameters $\log_{10} M_c$, $\theta_{\rm ej}$, $\eta_{\delta}$) producing the best independent fits at different redshifts. For the parameter $\log_{10}M_c$ (left-hand panel) a clear trend is visible, the values decreasing with increasing redshifts. The decrease is most pronounced for {\tt cosmo-OWLS 8.0}, {\tt cosmo-OWLS 8.5} and {\tt BAHAMAS}, while it is only week for {\tt OWLS}. For the parameter $\theta_{\rm ej}$ (middle panel), no clear and coherent trend is discernible. The values seem to oscillate and then increase, except for {\tt OWLS} where they slightly decrease towards higher redshifts. Regarding the third parameter $\eta_{\delta}$ (right-hand panel), a decreasing trend can be observed again. The slope of this decrease is similar between the simulations except for the cases where the parameter is pushed towards the edge of the allowed prior range (which is at $\eta_{\delta}=0.05$).

Motivated by the trends illustrated in Fig.~\ref{fig:3param_vs_z}, we allow the parameters to vary with redshift, following the simple relation 
\begin{eqnarray}
X(z) = X_0 (1+z)^{-\nu_X}.
\label{eq:param_zform}
\end{eqnarray}
Here $X$ stands for either $\log_{10} M_c$, $\theta_{\rm ej}$, or $\eta_{\delta}$. With this extended parametrisation, we have formally doubled the number of parameters, each former parameter now being described by its value at $z=0$ (given by $X_0$) and its power-law index (given by $\nu_X$) describing the redshift evolution. However, based on the trends visible in Fig.~\ref{fig:3param_vs_z} and in order to avoid an excessive growth of the parameter space, we fix the parameters $\nu_{\theta_{\rm ej}}$ and $\nu_{\eta_{\delta}}$ to 0 and 0.6, respectively. This means that we are left with a four-parameter model with $\log_{10}M_{c,0}$, $\theta_{\rm ej,0}$, $\eta_{\delta,0}$, and $\nu_{\log_{10}M_c}$ as freely variable parameters.



\begin{figure}
     \centering
     \begin{subfigure}[b]{0.95\textwidth}
         \centering
         \includegraphics[width=\textwidth]{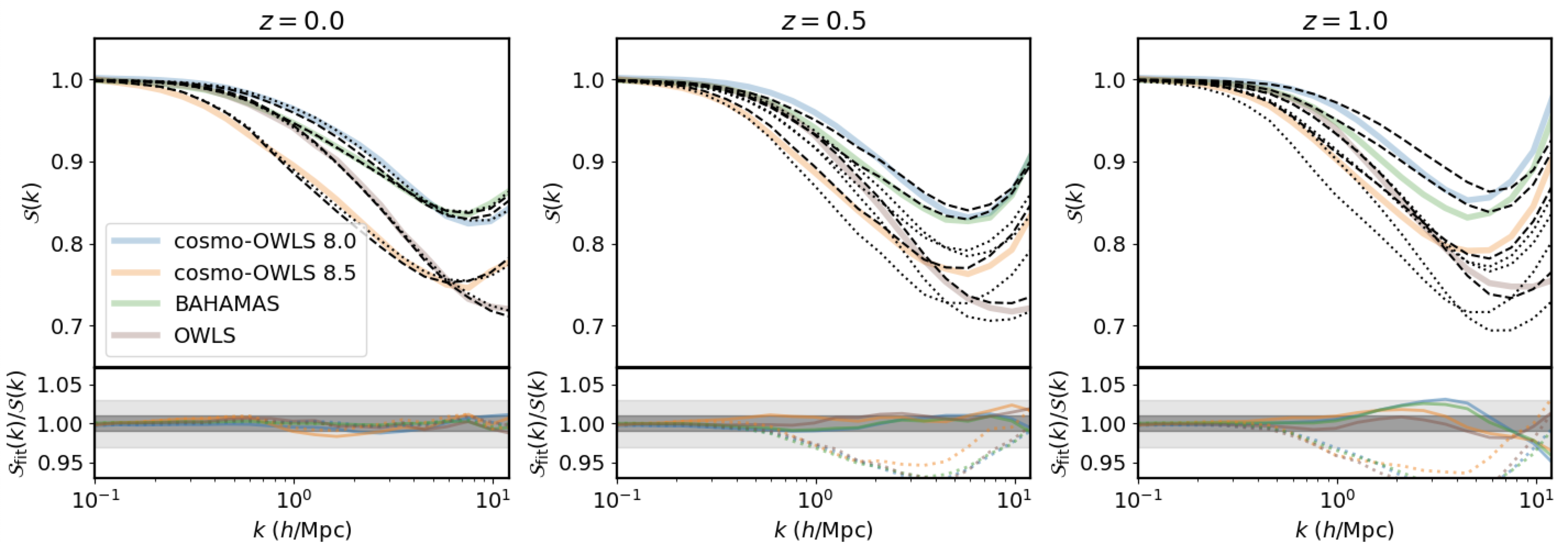}
         \label{fig:zdep_3param_0z1}
     \end{subfigure}
     \hfill
     \begin{subfigure}[b]{0.65\textwidth}
         \centering
         \includegraphics[width=\textwidth]{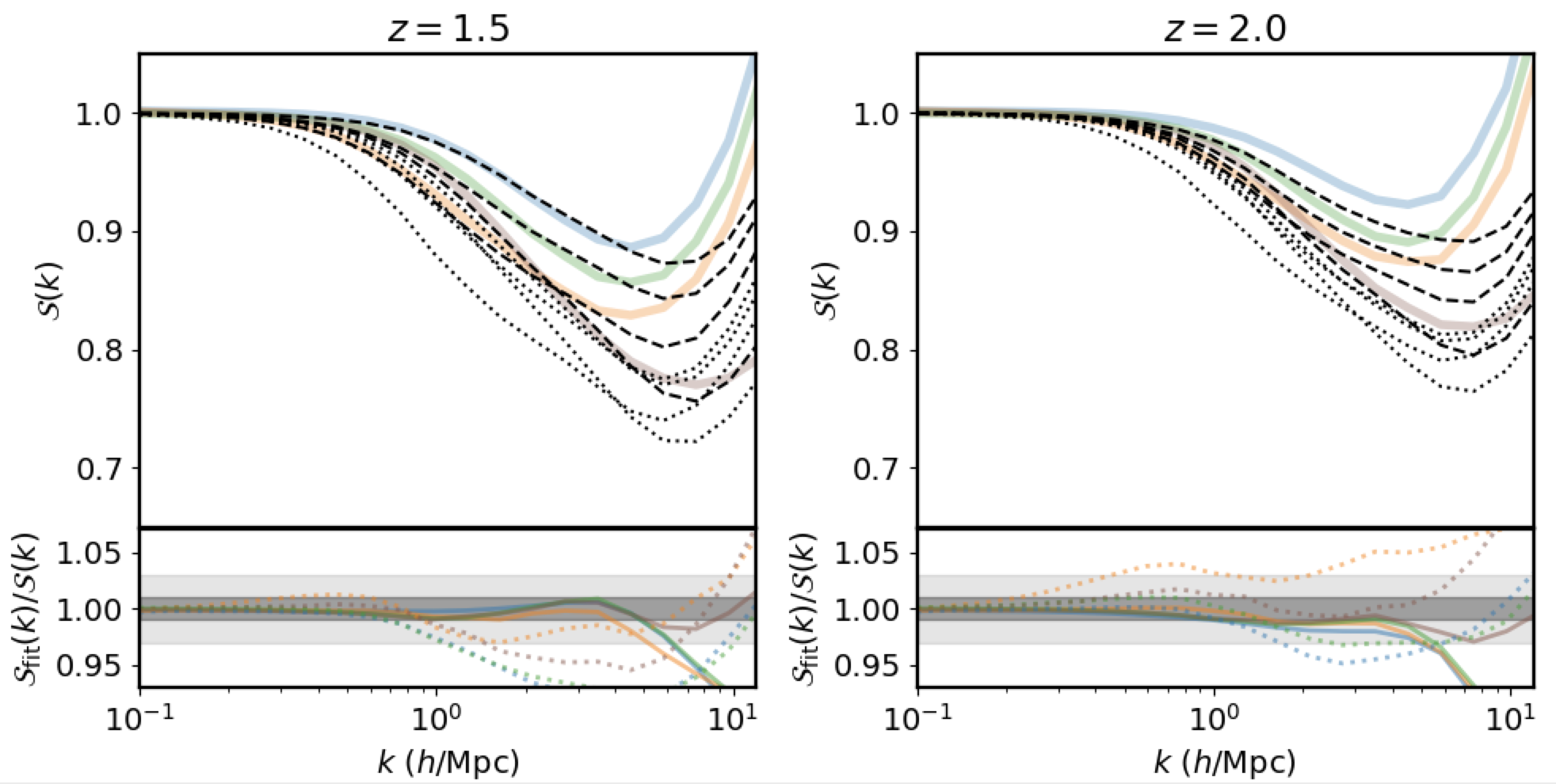}
         \label{fig:zdep_3param_1z2}
     \end{subfigure}
        \caption{Fits to the simulated baryonic power suppression ($\mathcal{S}$) assuming an explicit redshift dependence of the model. The coloured lines show the simulation results, while the dashed black lines correspond to the 4-parameter model which includes a power-law redshift dependence of the $\log_{10}M_c$ parameter (see text). Note that the model is fitted to the simulation results from all redshifts simultaneously. For comparison, we also show the results from the redshift-independent 3-parameters model fitted solely to the $z=0$ results.
        }
        \label{fig:best_fit_3param_zdep}
\end{figure}

In Fig.~\ref{fig:best_fit_3param_zdep}, we compare the new 4-parameter model (black dashed lines) to the results from simulations (coloured lines). This time, we do not fit the signal at different redshifts independently, but we perform a fit to all three redshift bins, allowing all four parameters ($\log_{10}M_{c,0}$, $\theta_{\rm ej,0}$, $\eta_{\delta,0}$, $\nu_{\log_{10}M_c}$) to vary simultaneously. We obtain a 3-percent agreement for all redshifts up to $k\sim5$ h/Mpc. For lower redshifts ($z\leq1$) the agreement is at the level of 3 percent up to $k\sim10$ h/Mpc.

Next to the 4-parameter model with redshift dependence, we also show the results for the redshift independent 3-parameter model as dotted black lines. This time, however (and unlike Fig.~\ref{fig:best_fit_3param}), the model parameters are only fitted to the simulation results at $z=0$. This means that the black dotted lines indicate the expected redshift dependence of $\mathcal{S}$ provided none of the model parameters are allowed to vary with redshift. A comparison between the dashed and dotted lines also illustrates the obtained gain when adding one additional parameter describing the explicit redshift dependence of $\log_{10} M_c$.

At the end of this section we want to emphasize that, while at least one redshift-dependent model parameter is clearly required to describe the simulation, there is currently no direct evidence from observations that model parameters have to vary with redshift. Indeed, Refs.~\cite{chiu2018baryon,sanders2018hydrostatic} find no significant redshift dependence in the X-ray cluster gas fraction between $z=0$ and 1. Note, however, that the uncertainties on these measurements remain rather large.

\section{Conclusions}
\label{sec:conclusion}

In this work we built an emulator for the power suppression signal $\mathcal{S}=P/P_{N- \rm body}$, which is primarily caused by baryonic feedback processes pushing gas out of galaxies into the intergalactic space. The baryonic effects are modeled using a slightly modified version of the baryonification method described in Ref.~\cite{Schneider2019QuantifyingCorrelation}. The baryonification method relies on parametrised halo profiles and is implemented as a post-processing step into gravity-only $N$-body simulations. While being many orders of magnitude faster than running full hydro-dynamical simulations, baryonification is not directly suitable for Bayesian inference using MCMC, as it involves reading and writing large $N$-body output files. Constructing an emulator based on a large training set of baryonified simulations is therefore a natural and efficient approach.

In a first step, we build five independent emulators at different redshifts ($z=0$, 0.5, 1.0, 1.5, 2.0) assuming eight free parameters. Five parameters are related to the gas profile ($\log M_c$, $\mu$, $\theta_{\rm ej}$, $\gamma$, $\delta$), two to the stellar fractions ($\eta$, $\eta_{\delta}$), and one parameter ($f_b=\Omega_b/\Omega_m$) captures effects from cosmology. The emulators are constructed using a modified kriging method, known as \emph{kernel partial least squares} (KPLS), which is a fast and memory-efficient emulation technique well suited for MCMC analyses. We show that the emulators have an accuracy of better than 1 percent at $z=0$, better than 2 percent at $z=0.5$, 1.0, 1.5, and better than 2.5 percent at $z=2$ for 95 percent of the cases and over all $k$-modes considered ($k\leq12.5$ h/Mpc).

The emulators (and the underlying baryionfication method) are validated using suites of hydro-dynamical simulations, such as {\tt cosmo-OWLS 8.0}, {\tt cosmo-OWLS 8.5}, {\tt BAHAMAS}, {\tt OWLS}, {\tt Illustris-TNG}, and {\tt Horizon-AGN}. We find that the emulators can fit all simulation results to better than one percent for $z=0$, 0.5, 1, and 1.5 over all scales ($k\leq12.5$ h/Mpc). Only for $z=2$, the agreement degrades somewhat but stays at the 3 percent level until $k\lesssim 5$ h/Mpc.


One of the important features of the baryonification model is that its parameters are connected to halo profiles and are therefore  physically meaningful. This makes it possible to constrain baryonic parameters using observational information from the gas and stellar distribution around clusters. We investigate this connection by testing the ability of the baryonification model to recover the power suppression signal from simulations when provided with information on the gas and stellar fraction of haloes alone. We show that the model is indeed capable to recover the power spectrum based solely on halo gas and stellar fractions of galaxy groups and clusters. Inversely, it can also predict the gas and stellar fractions based on the matter power spectrum alone. This important cross-check opens the door towards cross-correlation studies between e.g. gas and weak-lensing observations.

As a first step in that direction, we perform a case study where we use X-ray and optical cluster observations to predict the baryonic power suppression with the help of the baryonification model. We show that X-ray observations point towards a relatively strong baryonic suppression signal, exceeding the percent suppression at $k\sim0.1-0.4$ h/Mpc and growing to a maximum of more than 20 percent at around $k\sim 7$ h/Mpc. Such a suppression signal is in broad agreement with {\tt OWLS}, {\tt cosmo-OWLS}, and {\tt BAHAMAS}, but significantly stronger than the predictions from e.g. {\tt Illustris-TNG} and {\tt Horizon-AGN}.

In a further step, we reduce the amount of model parameters from seven to five, to three and to one, investigating the level of agreement that can be maintained between the simulations and the emulators. 
Intriguingly, we find the agreement to remain very good for the 5 and 3 parameter models, with deviations from simulations that stay well below 3 percent for most redshifts and $k$-ranges. Only the one parameter model performs significantly worse, with 3 percent agreement up to $k\sim 2$ h/Mpc and strong deviations at smaller scales.

Finally, we study the redshift dependencies of the baryonification parameters, that are required to reproduce the simulated power suppression signal over cosmic time. Using our three-parameter model, we find that we need at least one additional free parameter to describe the inherent redshift dependence of the aforementioned hydro-dynamical simulations. We thus provide a final four-parameter model that can reproduce simulation results at the level of a few percent for wave-modes below $k\sim10$ h/Mpc and redshifts below $z=2$. 

The baryonic emulator from this paper is available at \url{https://github.com/sambit-giri/BCemu} along with the documentation.

\acknowledgments

This work is
supported by the Swiss National Science Foundation via the grant \texttt{PCEFP2\_181157}. 
All the MCMC runs were done using the {\tt emcee} \cite{foreman2013emcee} package.
For analysis and plotting, we have used the following software packages: {\tt numpy} \cite{van2011numpy}, {\tt matplotlib} \cite{hunter2007matplotlib} and {\tt corner} \cite{corner}.
\RefReportTwo{We would like to thank the anonymous referee for helping us improve this work with their constructive comments.}

\appendix

\section{Construction of the training set}
\label{sec:const_training_set}

\RefReport{
\begin{figure}[t] 
 \centering
  \includegraphics[width=0.55\textwidth]{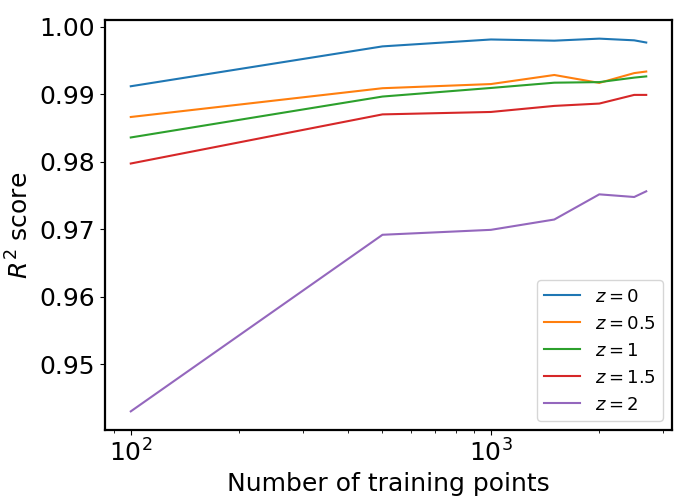}
 \caption{The $R^2$ score estimated for the testing set when emulator is built using using training set of various sizes. 
 }
 	\label{fig:accuracy_vs_nTaining}
\end{figure}

In this section, we describe our procedure for building the training set. We started by randomly sampling the 8-dimensional parameter space of the baryonification model with 100 points and subsequently appended the training set with batches of 100 points. In Fig.~\ref{fig:accuracy_vs_nTaining}, we show the accuracy ($R^2$ score) of the emulators as a function of the number of points in the training set. We find that all emulators attain a high accuracy with $\sim$1000 training points for $z<2$. We keep on increasing the training points (2700 points) until we achieve an $R^2$ score of 0.99 at $z=1.5$. With the final training set, we get an $R^2$ score of 0.975 at $z=2$.
}

\section{Method used for reducing the number of baryonic parameters}
\label{sec:model_selection_method}

Here we describe the procedure used to find the values of the parameters to fix in order to reduce the number of baryonification model parameters.
In this study, we fix two parameter at a time, going from the original 7 baryonic parameters to 5, 3, and 1.
For the first step (from seven to five parameters), we choose two out of the seven parameters and bin them within their ranges (see Table~\ref{table:range_values}) into 30 bins. After fixing these two parameters to a certain value, we fit this five parameter model to the $\mathcal{S}(k)$ from hydro-dynamical simulations. We sum the likelihood values of the best fitting model of all the simulations. Note that we use $\mathrm{S}(k)$ from multiple redshifts and we fit them independently.

\begin{figure}[t] 
 \centering
  \includegraphics[width=1.0\textwidth]{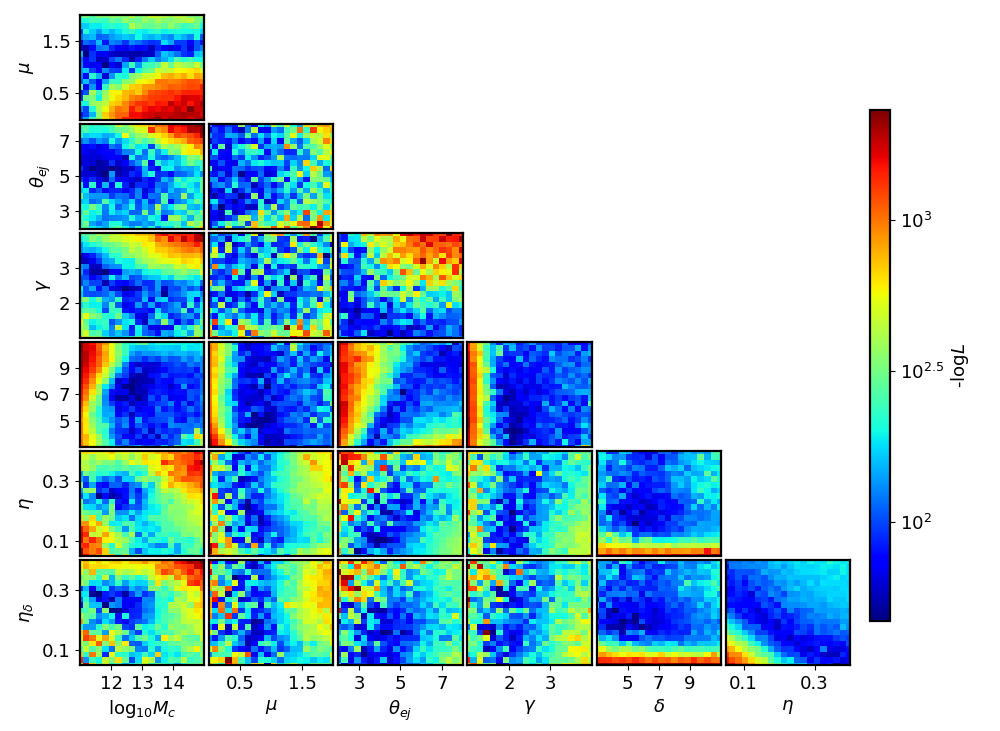}
 \caption{Best fitting 5-parameter models for different choices of fixed parameters. Each sub-panel shows a combination of fixed parameters with $-\log\mathcal{L}$ summed over all simulations and redshifts considered. Dark-blue (dark-red) areas correspond to 5-parameter models that provide the best (worst) global fit to the simulations. Based on this plot, we fix $\delta=7$ and $\eta=0.2$ for the 5-parameter model used in the main text.}
 	\label{fig:logL_5param_allsim}
 \end{figure}
 
\begin{figure}[t] 
 \centering
  \includegraphics[width=0.85\textwidth]{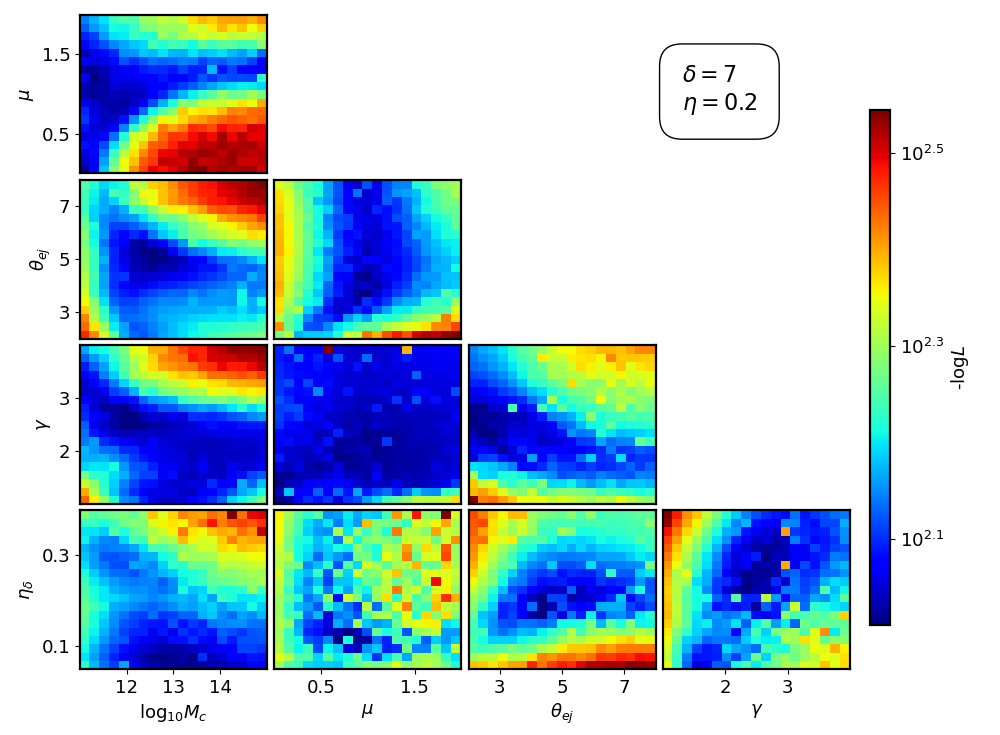}
 \caption{Best fitting 3-parameter models for different choices of fixed parameters (starting from the 5-parameter model selected above). Each sub-panel shows a combination of fixed parameters with $-\log\mathcal{L}$ summed over all simulations and redshifts. Dark-blue (dark-red) areas correspond to 3-parameter models that provide the best (worst) global fit to the simulations. Next to fixing $\delta=7$ and $\eta=0.2$ (see construction of the 5-parameter model), we fix $\mu=1$ and $\gamma=2.5$ to obtain the 3-parameter model used in the main text.}
 	\label{fig:logL_3param_allsim}
 \end{figure}
 
\begin{figure}[t] 
 \centering
  \includegraphics[width=0.55\textwidth]{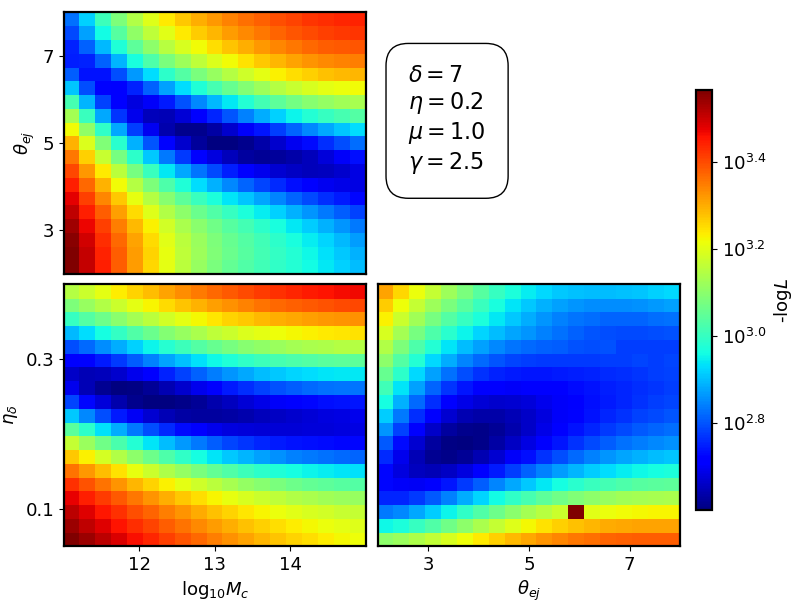}
 \caption{Best fitting 1-parameter models for different choices of fixed parameters (starting from the 3-parameter model selected above). Each sub-panel shows a combination of fixed parameters with $-\log\mathcal{L}$ summed over all simulations and redshifts. Dark-blue (dark-red) areas correspond to the 1-parameter models that provide the best (worst) global fit to the simulations. Next to fixing $\delta=7$, $\eta=0.2$, $\mu=1$ and $\gamma=2.5$ (see construction of the 3-parameter model), we fix $\theta_{\rm ej}=3.5$ and $\eta_{\delta}=0.2$ to obtain the 1-parameter model used in the main text.}
 	\label{fig:logL_1param_allsim}
\end{figure}

For the model selection, one can use the Akaike information criterion (AIC) and the Bayesian information criterion (BIC) that is given as, 
\begin{eqnarray}
 \mathrm{AIC} = 2N_p - 2\mathrm{log}\mathcal{L} \ ,~~~ \\
 \mathrm{BIC} = N_p\mathrm{log}(n) - 2\mathrm{log}\mathcal{L} \ .
 \label{eq:AIC_BIC}
\end{eqnarray}
In the above equations, $N_p$, $n$ and $\mathcal{L}$ are the number of parameters, number of observations fitted and likelihood value respectively. The \textit{Occam's razor} argument is built into these criteria, and models with minimised AIC and BIC are preferred. 
See Refs.~\cite{burnham2002practical,stoica2004model} to read more about model selection approaches. 

Fig.~\ref{fig:logL_5param_allsim} shows the negative sum of all the likelihood values of the best fit models to the hydro-dynamical simulations (for description of the simulations, see Sec.~\ref{sec:hydro_sims}). As $N_p$ and $n$ is the same for the models fitted, the AIC and BIC only depend on the log$\mathcal{L}$. In this figure, the dark-blue regions correspond to the best five-parameter models. 
Based on Fig.~\ref{fig:logL_5param_allsim}, we propose to fix $\delta$ and $\eta$ to 7 and 0.2 respectively.
In Fig.~\ref{fig:best_fit_5param}, we show the best fit models to the hydro-dynamical simulations using the five parameter representations.


As a next step, we further reduce the numbers of parameters from 5 to 3.
Fig.~\ref{fig:logL_3param_allsim} shows the negative sum of likelihood values ($-\log\mathcal{L}$) for all 3-parameter models with fixed pairs of parameters fitted to the simulations. Dark-blue areas correspond to models that provide the best global fits over all simulations and redshifts considered. 
Based on Fig.~\ref{fig:logL_3param_allsim}, we propose to fix $\mu$ and $\gamma$ to 1 and 2.5, respectively, ending up with the 3-parameter model used in the main text.

The final step consists of reducing the amount of free parameters from 3 to 1. In Fig.~\ref{fig:logL_1param_allsim}, we show the values of $-\log\mathcal{L}$ for all 1-parameter models fitted to the simulations. Dark-blue regions again correspond to the models that provide the best overall fit to the simulations. Based on the data shown in Fig.~\ref{fig:logL_1param_allsim}, we fix $\theta_\mathrm{ej}$ to 3.5 and $\eta_\delta$ to 0.2 to obtain the 1-parameter model used in the main text of this paper (see Sec.~\ref{sec:reduce_param}).
Note that, contrary to the reductions from 7 to 5 and 5 to 3 parameters, the AIC and BIC values are significantly larger for the 3 to 1 parameter reduction. 
\RefReport{We have listed the AIC and BIC values for the best fit models to all the simulations in Table~\ref{table:param_reduction_logL}.}
This suggests that the  3-parameter baryonification model consists of the best compromise between accuracy and small number of free parameters optimal for Bayesian inference analyses.


\begin{table}[h!]
\centering
\begin{tabular}{c c c} 
 Model & AIC & BIC  \\ 
 \hline \hline
 7 parameters & 28.3 & 34.1 \\ 
 5 parameters & 88.4 & 92.6 \\ 
 3 parameters & 110.6 & 113.1 \\ 
 1 parameter  & 838.7 & 839.6 \\ 
\end{tabular}
\caption{\RefReport{List of the AIC and BIC values from the best fit models to all the hydro-dynamical simulations.}}
\label{table:param_reduction_logL}
\end{table}

\section{Single parameter baryonification model}
\label{sec:single_BCM}

 \begin{figure}[t] 
 \centering
  \includegraphics[width=1.0\textwidth]{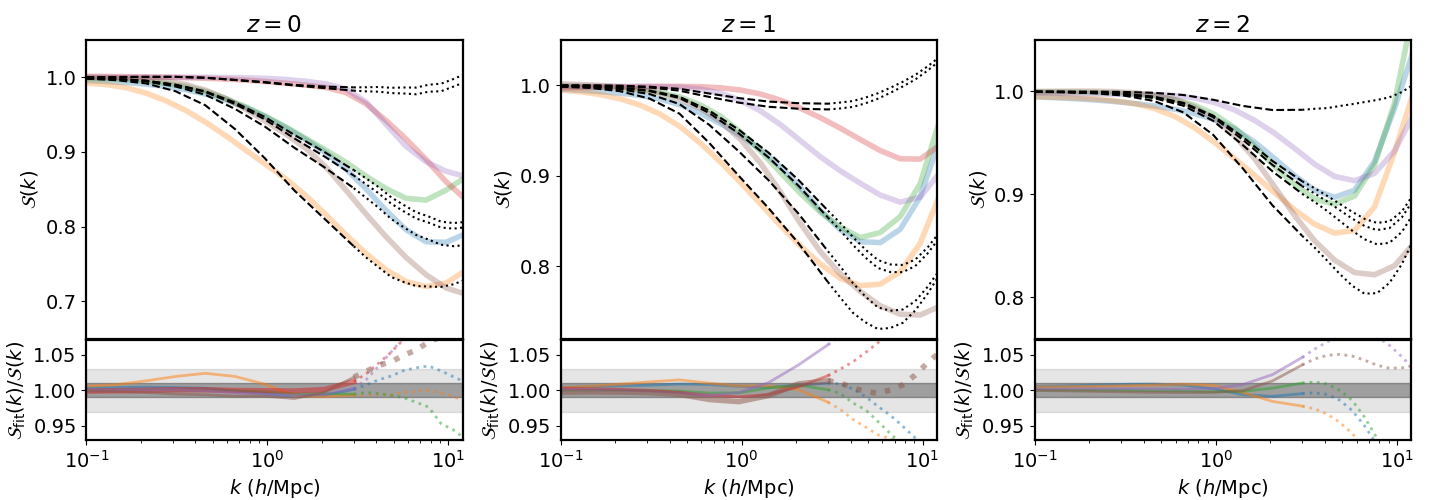}
 \caption{Fits to the simulated baryonic power suppression signal ($\mathcal{S}$) at large scales ($k\lesssim 3$ h/Mpc) for one-parameter baryonification model. From left to right we show results from redshift 0, 1, and 2 (which are fitted independently). Coloured lines correspond to simulation results, dashed black lines to the emulator. We only fit up to $k=3$ h/Mpc, beyond this range the emulator results are shown as dotted lines.}
 	\label{fig:best_fit_1param_k_large}
 \end{figure}
 
Our one-parameter baryonification model struggles to reproduce the baryonic power suppression ($\mathcal{S}$) from simulations when fitted to scales between $k=0.1$ h/Mpc and 12.5 h/Mpc (see Fig.~\ref{fig:best_fit_1param} and description in Sec.~\ref{sec:reduce_param}). 
In this appendix, we show that $\mathcal{S}$ can be reproduced to much better accuracy when only fitted to large scales, i.e. until $k$-modes of $k=3$ h/Mpc. 

In Fig.~\ref{fig:best_fit_1param_k_large}, we show the fits to the power spectra suppression for $k\leq3$ h/Mpc. The three panels show the results at redshift 0, 1, and 2. When focusing on large scales alone (with $k\lesssim 2$ h/Mpc), the agreement between our model and the simulations stays within 3 percent for all redshifts.
At smaller scales (larger $k$-modes) the curves diverge from the simulation results, leading to substantial errors of up to 10-20 percent at $k\sim 10$ h/Mpc.


\bibliographystyle{JHEP}
\bibliography{Mendeley,reference}
\end{document}